\documentclass[10pt,showpacs,preprintnumbers,twocolumn]{revtex4}

\usepackage{times}
\usepackage{amsmath}
\usepackage{amssymb}
\usepackage{graphicx}
\usepackage{dcolumn}
\usepackage{bm}

\newcommand{\diagmat}{\mathop{\rm diag}}
 \newcommand{\nand}{{\,\barwedge\,}}      
 \newcommand{\bzero}{{\bf 0}}
 \newcommand{\bone}{{\bf 1}}
 \newcommand{\rT}{{\mathrm T}}
  
  \newcommand{\rB}{\mathrm{B}}

   \newcommand{\re}{{\rm e}}

   \newcommand{\face}{{\rm face}}
    \newcommand{\edge}{{\rm edge}}
   \newcommand{\site}{{\rm site}}
   \newcommand{\diag}{{\rm diag}}

   \newcommand{\FDOI}{{\rm FDOI}}

 \newcommand{\x}{{\times}}      
 \newcommand{\mZ}{{\mathbb Z}}  
 \newcommand{\bP}{\mathbf{P}}  
 \newcommand{\bH}{\mathbf{H}}  
 \newcommand{\bI}{\mathbf{I}}  
 \newcommand{\bh}{{\mathbf h}}  
 \newcommand{\bU}{{\bf U}}  
 \newcommand{\bG}{{\mathbf G}}  
 \newcommand{\bg}{{\mathbf g}}  

 \newcommand{\cS}{{\mathcal S}} 
 \newcommand{\cD}{{\mathcal D}} 
  \newcommand{\cU}{{\mathcal U}} 
 \newcommand{\cN}{{\mathcal N}} 
 \newcommand{\cI}{{\mathcal I}} 

 \newcommand{\Ups}{{\Upsilon}}  
 \newcommand\op{\oplus}
  
 \newcommand\sS{{\sf S}}        
 \newcommand\sE{{\sf E}}        
 \newcommand\sF{{\sf F}}        
 \newcommand\sD{{\sf D}}        

  \newcommand\sU{{\sf U}}        
  \newcommand\sA{{\sf A}}        
  
\newcommand\bra{{\langle}}
\newcommand\ket{{\rangle}}

\newcommand{\cL}{{\cal L}}

\def\<{\leqslant}           
\def\>{\geqslant}           

\def\d{\partial}
\def\wh{\widehat}
\def\wt{\widetilde}

\input{epsf}

\begin{document}


\title{\Large Face-to-diagonal reduction of Kramers-Wannier approximation for cubic lattice particle systems with nearest
    neighbour exclusion
}
\author{Igor G. Vladimirov}

\email{igor.g.vladimirov@gmail.com}

\affiliation{UNSW Canberra, ACT 2600, Australia}

\begin{abstract}
The paper is concerned with interacting particle systems on the simple cubic lattice obeying the nearest neighbour exclusion (NNE). This constraint forbids any two neighbouring sites of the lattice to be simultaneously occupied, thus reducing the set of admissible configurations for the cubic cell and its subclusters such as edges and faces. This reduction extends applicability of Kikuchi's Cluster Variation Method (CVM) with higher-order clusters to systems with complex site configurations and short-range ordering, which would be impractical beyond the NNE framework because of the ``curse of dimensionality''. For edges of the cubic cell, which are the operational clusters of the Bethe-Peierls  entropy approximation in the CVM hierarchy, the edge-to-site reduction of the entropy cumulants was studied previously. In extending the earlier results, we develop a face-to-diagonal reduction of the Kramers-Wannier entropy approximation of the CVM in the NNE setting. We also outline an application of the resulting approximation to thermodynamic modeling of disordered condensed media, such as liquid silicates, and discuss combinatorial and numerical aspects of the implementation of this approach.
\end{abstract}

\maketitle

\section{Introduction}
\label{sec:intro}


The present study continues the theme of \cite{Vladimirov} and is concerned
with the statistical mechanical modeling of classical
interacting particle systems on the simple cubic lattice with
nearest neighbour exclusion (NNE). Assuming the site configurational
space to be a finite set, and distinguishing one of its elements as \emph{vacancy} while interpreting the others as \emph{occupied} site
configurations, NNE forbids any two nearest sites of the lattice to be simultaneously occupied.
With the number of site configurations not being limited to two, NNE
generalizes a similar constraint which is used in the hard-core lattice
liquid/gas models \cite{Verhagen,Baxter}. Moreover, it provides a more
flexible setting for the modeling of \emph{disordered} condensed media
in comparison with the face-centered-cubic (FCC) lattice formulation.

For NNE-constrained cubic lattice particle systems, we develop a \emph{face-to-diagonal reduction} of the Kramers-Wannier entropy
approximation~\cite{Huang}. The latter constitutes the
second level in the hierarchy of the Cluster Variation Method (CVM)
approximations~\cite{An, Kikuchi1, Kikuchi2, Kikuchi3, Kikuchi_2002,
Moran, Pelizzola} and takes into account the equilibrium statistical
correlations in the particle system within faces of cubic cells.
Accordingly, the configurational entropy of the system per lattice site, which is known to be intractable
in three dimensions, is approximated by a
linear combination of the site, edge and face entropies
weighted by Kikuchi-Barker coefficients for
the simple cubic lattice. Thus, the faces of cubic cells of the lattice play the role of basic
clusters with their edges and sites as subclusters.

The NNE constraint allows the Kramers-Wannier entropy approximation to be reduced to the site and face-diagonal entropies, where the latter are associated with  face diagonals of cubic cells. More precisely, the resulting entropy approximation is organized as a linear combination of the site and face-diagonal entropies along with the Shannon mutual information \cite[p.~19--22]{Cover} between the occupancies of the  nearest neighbours and the face diagonals, which are induced by the NNE constraint. That is, the entropy approximation is essentially reduced to one- and two-site clusters. The present study takes advantage of this dimensionality reduction potential of the NNE setting in order to advance practical applicability of the Kramers-Wannier approximation towards lattice models of disordered condensed media with complex site configurations and short-range ordering.

We apply the NNE-induced face-to-diagonal reduction of the Kramers-Wannier
approximation of the \emph{configurational} entropy to a
class of chemical systems,  where site configurations represent
spatial arrangements of \emph{coordination entities} consisting of a
\emph{central atom} at a site of the simple cubic lattice and a surrounding
array of ligands. The locations of the central atoms are
subjected to NNE, whilst the ligands are allowed to reside at sites of an
interstitial lattice and may be shared by several ``overlapping''
coordination entities. The overlap induces additional geometric
constraints which single out a class of admissible pairs of coordination entities
centered at the end sites of face diagonals.

The energetics of the system is parameterized by interaction energies which are ascribed to representatives of the isotropy equivalence classes into which the admissible face-diagonal configurations are partitioned by the action of the full octahedral symmetry group of 48 isometries of the simple cubic lattice~\cite{Cotton}. We develop a theory which allows the equilibrium Gibbs energy and related thermodynamic quantities to be approximately computed for given values of the energy parameters. This approach  is based on minimizing the variational free energy density approximation (per central atom)  over admissible face-diagonal and site probability mass functions (PMFs) subject to the compatibility (marginalization) and balance relations.

The constrained minimization problem is solved by using a
separation-of-variables technique, not dissimilar to that in Dynamic
Programming. More precisely, the optimization problem is decomposed into a
family of pairs of optimization problems which share a
common scalar parameter but are  solved separately, with the dependencies on the pressure and
chemical composition of the system entering these two
problems in isolated ways. The solution of the first problem is
reduced to finding a root of a decic polynomial, whilst the second problem
resembles the Bethe-Peierls approximation for an FCC lattice, though
with different coefficients in the entropy cumulants. These solutions
are then ``assimilated'' by minimization over the master parameter,  which leads to a complicated (non-additive)  dependence on the pressure and chemical composition. This computational
approach is applicable to a wide range of pressure values (in principle, including tectonic pressures).

In order to develop the NNE-constrained entropy approximation and the solution of the optimization problem, we employ the probabilistic concepts such as the above mentioned Shannon information and conditional entropy~\cite[p.~17--20]{Cover} (which are used here similarly to \cite{Vladimirov}) together with Boolean random variables and logical operations on them,  including the Sheffer stroke~\cite[p.~51]{Enderton}.  This combination provides an efficient machinery for entropy theoretic computations under geometric constraints.

The face-to-diagonal reduction of the Kramers-Wannier approximation is then applied to thermodynamic modeling  of a binary liquid silicate ${\rm SiO_2}$--${\rm M_2O}$ formed from silica and the oxide of a univalent metal M, and the combinatorial and numerical aspects of this application are discussed. Using the structural model from Section~IV of \cite{Vladimirov}, which relies on the qualitative insights into the internal structure of silicate melts~\cite{Gaskell,Mysen}, the liquid silicate is modelled as an assemblage of Si--O--Si, Si--O--M and M--O--M second nearest neighbour bonds (SNNBs) centered at oxygen atoms residing at sites of the simple cubic lattice subject to NNE.

Since the present study takes into account face-diagonal correlations, it also leads to a refined internal energy model which, as mentioned above, sums the energies of interaction between SNNBs at the end sites of face diagonals of cubic cells. This includes, as a particular case, the approach of the existing thermodynamic models to liquid silicates~\cite{Gaskell, Mysen}, such as the Quasi-Chemical Model and its modifications~\cite{Blander,Pelton1,Pelton2}, where the internal energy is assumed to be composed of SNNB energies.

Practical model calibration for specific systems, such as sodium silicate ${\rm SiO_2}$--${\rm Na_2O}$, depends on availability of  an efficient numerical algorithm for solving a particular non-convex minimization problem. This requires an additional research into the implementation of the model (see Section~\ref{sec:Lagrange} for details) and is beyond the scope of the present paper.

The organization of the paper is as follows.  Section~\ref{sec:setting} describes the NNE-constrained cubic lattice particle systems being considered. These are instantiated in Section~\ref{sec:occupied_site_confs}, which specifies the set of occupied site configurations and related balance equations for chemical systems with short-range ordering, such as liquid silicates. Section~\ref{sec:edge-to-site} revisits the edge-to-site entropy reduction from \cite{Vladimirov} for completeness. Section~\ref{sec:face-to-diagonal} expresses the face entropies in terms of the appropriate face-diagonal entropies. Sections~\ref{sec:face_PMF_equations} and \ref{sec:admissibility} describe marginalization and other constraints for the face-diagonal PMFs. Section~\ref{sec:face-to-diagonal_KW} carries out the face-to-diagonal reduction of the Kramers-Wannier entropy density approximation under NNE. In combination with the internal energy model of Section~\ref{sec:internal_energy}, the resulting entropy density estimate is used in Section~\ref{sec:Gibbs_energy_estimate} in order to formulate the approximate computation of the Gibbs energy through minimizing the variational free energy estimate per central atom over admissible face-diagonal and site PMFs subject to the marginalization and balance constraints. The solution of this constrained optimization problem is considered in Sections~\ref{sec:conditioned}--\ref{sec:separation} under an additional isotropy assumption of Section~\ref{sec:isotropy} which further reduces the problem dimensionality  by an order of magnitude. To this end, Section~\ref{sec:separation} develops the separation-of-variables technique based on the conditioned representation of the site and face-diagonal entropies and the internal energy density from Section~\ref{sec:conditioned} along with isotropic versions of balance and marginalization equations from Sections~\ref{sec:iso_balance} and~\ref{sec:iso_marginalization}. Section~\ref{sec:binary_silicate} outlines an application of the statistical mechanical approach to thermodynamic modeling of a binary liquid silicate. Concluding remarks are given in Section~\ref{sec:conclusion}. Appendices provide subsidiary material.

\section{NNE-constrained cubic lattice setting}
\label{sec:setting}

We consider an interacting particle system at thermodynamic equilibrium on the simple cubic lattice $\mZ^3$
with a finite \emph{site configurational space}
\begin{equation}
\label{Omega}
    \Omega
    :=
    \{0\}
    \bigsqcup
    W.
\end{equation}
Here,  $0$ is interpreted as \emph{vacancy}, $W$ is a set of configurations for an \emph{occupied} site of the lattice (so that $0\not\in W$), and $\bigsqcup$ denotes the union of disjoint sets.
The equilibrium spatial arrangement of the particle system is described by an $\Omega$-valued \emph{homogeneous} random field $\xi := (\xi_z)_{z \in \mZ^3}$, where $\xi_z$ is the state of site $z \in \mZ^3$. The homogeneity of $\xi$ is understood in the usual sense as the invariance of its multi-point probability distributions with respect to  translations of the lattice \cite{Kindermann, Preston}.
\begin{figure}[htb]
\centering
\includegraphics[width=3.5cm]{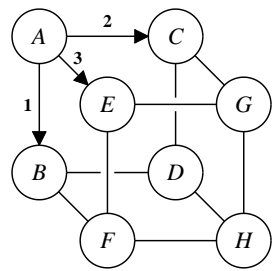}
\caption{    The states $A, \ldots, H$ of sites of the cubic cell.
    The nearest neighbours are connected by edges.
    The arrows represent the Cartesian coordinate axes.
}
\label{fig1}
\end{figure}
The states of sites of the cubic cell of the lattice are denoted by $A, B, C, D, E, F, G, H$ as shown in Fig.~\ref{fig1}.  These are identically distributed random variables with values in the set $\Omega$. Their common marginal probability mass function (PMF) $\sS := (\sS_u)_{u \in \Omega}\in [0,1]^{\Omega}$, which we will refer to as the \emph{site PMF}, is defined by
\begin{equation}
\label{S}
    \sS_u
    :=
    \bP(A = u),
    \qquad
    u \in \Omega,
\end{equation}
where $\bP(\cdot)$ is the underlying probability measure. Assuming that the equilibrium
random field $\xi$ is spatially ergodic, $\sS_u$ represents the relative
fraction of those sites in a macroscopically large (for simplicity, cubic)
fragment of the lattice which are at a given configuration $u \in \Omega$:
\begin{equation}
\label{Slim}
    \lim_{N\to +\infty}
    \frac{\#\{z \in Q_N:\ \xi_z = u \}}{N^3} = \sS_u,
\end{equation}
where the convergence holds almost surely. Here,
$\#(\cdot)$ is the number of elements in a finite set, and
\begin{equation}
\label{QN}
    Q_N := \{ 0, \ldots, N-1 \}^3
\end{equation}
is a discrete cube consisting of $N^3$ sites of $\mZ^3$.  The particular location of the cube does not affect the limit in (\ref{Slim}) in view of the homogeneity of the random field $\xi$.

For what follows, we assume that the particle system is constrained by the \emph{nearest
neighbour exclusion} (NNE) which forbids any two neighbouring sites of
$\mZ^3$ to be simultaneously occupied:
\begin{align}
\nonumber
    \bP(
        A\ne 0,
        B \ne 0
    )
    & =
    \bP(
        A\ne 0,
        C \ne 0
    )\\
\nonumber
     & =
    \bP(
        A\ne 0,
        E \ne 0
    )\\
\label{NNE_AB_AC_AE}
     & =
    0.
\end{align}
In view of the NNE constraint, the probability that two nearest sites of the lattice are both vacant is then computed by the inclusion-exclusion principle as
\begin{align}
\nonumber
    \bP(A=B=0)
    = &
    \bP(A=0) + \bP(B=0)\\
\nonumber
      & +
    \bP(A\ne 0, B\ne 0)-1\\
\label{00}
     = &    2\sS_0-1
\end{align}
(see also Eq.~(4) of
\cite{Vladimirov}),
where $\sS_0$ is the vacancy probability in accordance with (\ref{S}).
 Therefore, $\sS_0 \> \frac{1}{2}$, with the extreme value $\sS_0 =
\frac{1}{2}$ corresponding to the densest packing in $\mZ^3$, where one of any two nearest
sites of the lattice is occupied while the other is vacant \cite{Conway}.

\section{Occupied site configurations}
    \label{sec:occupied_site_confs}


Although the subsequent discussions are applicable to more general disordered condensed media, the interpretation of occupied site configurations in (\ref{Omega}) will be aimed at a particular class of chemical systems. More precisely, we interpret each $w\in W$ as a spatial arrangement of a \emph{coordination entity} which consists of a \emph{central atom}, occupying a site of the simple cubic lattice, and an array of ligands bound to it. The ligands are not necessarily accommodated by $\mZ^3$ and may reside at sites of  an interstitial lattice. However, their relative positions with respect to the central atom are specified for each occupied site configuration. Furthermore, the ligands are allowed to be shared by several ``overlapping'' coordination entities.

Let $n$ denote the number of constituent particle species in the system. Within each of the species,  particles are identical. For every $i = 1,\ldots, n$, the
$i$th species is endowed with a \emph{coordination number} $\nu_i$ in the sense
that any representative of the species is always shared by $\nu_i$ coordination
entities associated with distinct sites of the simple cubic lattice.

We use the convention that the 1st species, which is further referred to as
the \emph{central species}, is represented only by central atoms and
has coordination number $\nu_1 = 1$. More precisely, for each
occupied site configuration $w \in W$, the corresponding
coordination entity contains a single representative of the 1st
species and the particle is   the central atom in the entity.

For example, in the structural model of a binary silicate melt
${\rm SiO_2-M_2O}$ described in Section~IV of
\cite{Vladimirov}, where ${\rm M}$ is a univalent metal, the elements of the set $W$
represent $79$ second nearest neighbour bonds (SNNBs) ${\rm
X-O-Y}$. This model involves $n = 3$ species, with oxygen being the central species.
Considering that ${\rm Si}$ and ${\rm M}$ are the 2nd and 3rd particle species, their
coordination numbers are $\nu_2 = 4$ and $\nu_3 = 1$, respectively.

In view of the spatial ergodicity hypothesis of Section~\ref{sec:setting}, the numbers of particles of different species in a macroscopically large fragment of the lattice, consisting of $\cN$ sites, are asymptotically given by
\begin{equation}
\label{Ni}
    \cN_i
    \sim
    \frac{\cN}{\nu_i}
    \sum_{w\in W}
    \Gamma_{iw}
    \sS_w,
    \quad
    {\rm as}\
    \cN\to +\infty,
    \quad
    i =1, \ldots, n.
\end{equation}
Here, $\Gamma_{iw}$ denotes the number of particles of the $i$th species
in a site configuration $w$, and the denominator $\nu_i$ comes from the fact that particles of the $i$th species are counted $\nu_i$ times according to their coordination number. In particular,  the number of particles of the central species in the lattice fragment
is asymptotically given by
\begin{equation}
\label{N1}
    \cN_1
    \sim
    (1-\sS_0)
    \cN,
\end{equation}
where the conventions $\nu_1 = 1$ and $\Gamma_{1w} = 1$
for all $w \in W$ are used in combination with the probability that a lattice site is occupied:
\begin{equation}
\label{SS}
    \bP(A\ne 0)
    =
    1-\sS_0
    =
    \sum_{w\in W}
    \sS_w.
\end{equation}
The relative mole fractions of the constituent particle species with
reference to the central species can therefore be defined by
\begin{equation}
\label{Ni/N1}
    y_i
    :=
    \frac{\cN_i}{\cN_1},
\end{equation}
so that $y_1 = 1$. In what follows, the quantity $y_i$ will be referred to as
the \emph{centralized} mole fraction of the $i$th particle species.
By dividing both parts of (\ref{Ni}) by those of (\ref{N1}), it
follows that (\ref{Ni/N1}) is equivalent to
\begin{equation}
\label{cond_balance}
    \frac{1}{\nu_i}
    \sum_{w\in W}
    \Gamma_{iw}
    \wt{\sS}_w
    =
    y_i,
    \qquad
    i = 1, \ldots, n.
\end{equation}
Here, $\wt{\sS}_w$ denotes the conditional probability that a lattice site is at a state $w\in W$, provided this site is occupied. In accordance with (\ref{SS}) these conditional probabilities are computed as
\begin{equation}
\label{cond_site_PMF}
    \wt{\sS}_w
    :=
    \bP(A = w \mid A \ne 0)
    =
    \frac{\sS_w}{1-\sS_0},
    \qquad
    w \in W,
\end{equation}
and comprise the conditional PMF  $\wt{\sS} := (\wt{\sS}_w)_{w\in W} \in [0,1]^W$ for the state of an occupied site of $\mZ^3$.

\section{Edge-to-site entropy reduction}
\label{sec:edge-to-site}

In view of the NNE constraint (\ref{NNE_AB_AC_AE}), the common \emph{edge PMF} $\sE := (\sE_{uv})_{u,v\in \Omega} \in [0,1]^{\Omega^2}$ of the pairs $(A,B)$, $(A,C)$, $(A,E)$ (which are regarded as $\Omega^2$-valued random variables) is given by
\begin{align}
\nonumber
    \sE_{uv}
    := &
    \bP(A=u,B=v)\\
\nonumber
    = &
    \bP(A=u,C=v)\\
\nonumber
     =&
    \bP(A=u,E=v)\\
\label{edge_PMF}
    =&
    \left\{
    \begin{array}{ll}
    2\sS_0 - 1 & {\rm if}\ u = v = 0\\
    \sS_u & {\rm if}\ u \in W,\ v = 0\\
    \sS_v & {\rm if}\ u = 0 ,\ v \in W\\
    0 & {\rm if}\ u,v \in W
    \end{array}
    \right.,
\end{align}
where use is also made of (\ref{00}).
As discussed in \cite{Vladimirov}, the property that the edge PMF $\sE$ is
specified completely by the site PMF $\sS$  is an important  consequence  of the NNE constraint \footnote{In general, the joint probability distribution of two or more random variables is not specified uniquely by their marginal distributions.} which allows the common \emph{edge entropy}
\begin{equation}
\label{Hedge}
    \bH_{\edge}
     :=
    -
    \sum_{u,v \in \Omega}
    \Lambda(\sE_{uv})
\end{equation}
to be reduced to the site entropy
\begin{equation}
\label{Hsite}
    \bH_{\site}
    :=
    -
    \sum_{u \in \Omega}
    \Lambda(\sS_u)
\end{equation}
as
\begin{equation}
\label{Hedgesite}
    \bH_{\edge}
    =
    2\bH_{\site}
    -
    \varphi(\sS_0),
\end{equation}
see also Eq.~(9) of \cite{Vladimirov}. Here, use is made of  the shorthand notation
\begin{equation}
\label{Lambda}
    \Lambda(p) := p\ln p,
\end{equation}
with the standard convention $\Lambda(0) := 0$, and the
function $\varphi$ is defined by
\begin{equation}
\label{varphi}
    \varphi(p)
    :=
    \Lambda(2p-1) - 2\Lambda(p)
    =
    \Phi(p, p),
\end{equation}
where
\begin{equation}
\label{Phi}
    \Phi(p,q)
    :=
    \Lambda(p+q-1)
    -
    \Lambda(p)
    -
    \Lambda(q).
\end{equation}
The graph of $\varphi$ can be found in Fig.~2 of \cite{Vladimirov}. The
quantity $\Phi(p,q)$ coincides with the Shannon mutual information
\cite[p.~19--22]{Cover}
$$
    \bI(\alpha; \beta)
    :=
    \bH(\alpha)
    +
    \bH(\beta)
    -
    \bH(\alpha, \beta)
$$
between Boolean random variables $\alpha$ and $\beta$, which take
values in the set $\{0, 1\}$ with probabilities $p := \bP(\alpha=0) = 1-\bP(\alpha=1)$ and $q := \bP(\beta =
0) = 1-\bP(\beta=1)$ and are mutually exclusive in the sense that their Sheffer
stroke $\alpha \nand \beta:= \neg(\alpha \wedge \beta)$ (that is, the negated conjunction)
takes the truth value almost surely:
$\bP(\alpha \nand \beta = 1) =1$. The latter condition implies that $p+q\>
1$.

Hence, by denoting the indicator function of the set $W$ of
occupied site configurations in (\ref{Omega}) by $\cI_W(\cdot)$, the quantity
$\varphi(\sS_0)$ in (\ref{varphi}) can be represented in terms of the
Shannon information as
\begin{align*}
    \varphi(\sS_0)
    & =
    \bI(\cI_W(A); \cI_W(B))\\
    & =
    \bI(\cI_W(A); \cI_W(C))\\
    & =
    \bI(\cI_W(A); \cI_W(E)),
\end{align*}
where the Boolean random variables $\cI_W(A), \ldots, \cI_W(H)$ are the {\it
site occupancy indicators} (SOIs) introduced by Eq.~(24) of \cite{Vladimirov}.

Furthermore, the \emph{edge-to-site entropy reduction}, described by
(\ref{Hedgesite}), implies that $\varphi(\sS_0)$ is also the Shannon
information
\begin{align*}
    \varphi(\sS_0)
    & =
    2\bH_{\site}
    -
    \bH_{\edge}\\
    & =
    \bI(A;B)
    =
    \bI(A;C)
     =
    \bI(A;E)
\end{align*}
between the states of neighbouring lattice sites which is induced by the NNE constraint. Indeed, if the states of such sites were independent random variables, the edge entropy would be equal to $2\bH_{\site}$.

\section{Face-to-diagonal entropy reduction}
\label{sec:face-to-diagonal}

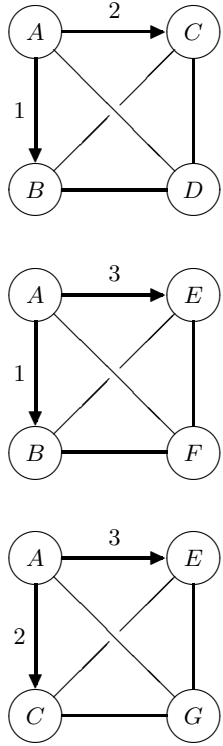
\begin{figure}[htb]
\begin{center}
\unitlength = 0.7mm

\linethickness{1.0pt}
\begin{picture}(50.00,160.00)


\put(10,150){\circle{10}}

\put(10,150){\makebox(0,0)[cc]{$A$}}

\put(10,145){\line(0,-1){19}}

\put(10,125.2){\makebox(0,0)[cb]{$\blacktriangledown$}}

\put(7,135){\makebox(0,0)[cc]{$1$}}

\put(25,154){\makebox(0,0)[cc]{$2$}}

\put(15,150){\line(1,0){19}}

\put(35,150){\makebox(0,0)[rc]{$\blacktriangleright$}}

\put(40,145){\line(0,-1){20}}

\put(15,120){\line(1,0){20}}

\put(13.5,146.5){\line(1,-1){23}}

\put(13.5,123.5){\line(1,1){10.5}}

\put(26.1,136.1){\line(1,1){10.4}}


\put(10,120){\circle{10}}

\put(10,120){\makebox(0,0)[cc]{$B$}}


\put(40,150){\circle{10}}

\put(40,150){\makebox(0,0)[cc]{$C$}}

\put(40,120){\circle{10}}

\put(40,120){\makebox(0,0)[cc]{$D$}}

\put(10,100){\circle{10}}

\put(10,100){\makebox(0,0)[cc]{$A$}}

\put(10,95){\line(0,-1){19}}

\put(10,75.2){\makebox(0,0)[cb]{$\blacktriangledown$}}

\put(7,85){\makebox(0,0)[cc]{$1$}}

\put(25,104){\makebox(0,0)[cc]{$3$}}

\put(15,100){\line(1,0){19}}

\put(35,100){\makebox(0,0)[rc]{$\blacktriangleright$}}

\put(40,95){\line(0,-1){20}}

\put(15,70){\line(1,0){20}}

\put(13.5,96.5){\line(1,-1){23}}

\put(13.5,73.5){\line(1,1){10.5}}

\put(26.1,86.1){\line(1,1){10.4}}


\put(10,70){\circle{10}}

\put(10,70){\makebox(0,0)[cc]{$B$}}


\put(40,100){\circle{10}}

\put(40,100){\makebox(0,0)[cc]{$E$}}

\put(40,70){\circle{10}}

\put(40,70){\makebox(0,0)[cc]{$F$}}


\put(10,50){\circle{10}}

\put(10,50){\makebox(0,0)[cc]{$A$}}

\put(10,45){\line(0,-1){19}}

\put(10,25.2){\makebox(0,0)[cb]{$\blacktriangledown$}}

\put(7,35){\makebox(0,0)[cc]{$2$}}

\put(25,54){\makebox(0,0)[cc]{$3$}}

\put(15,50){\line(1,0){19}}

\put(35,50){\makebox(0,0)[rc]{$\blacktriangleright$}}

\put(40,45){\line(0,-1){20}}

\put(15,20){\line(1,0){20}}

\put(13.5,46.5){\line(1,-1){23}}

\put(13.5,23.5){\line(1,1){10.5}}

\put(26.1,36.1){\line(1,1){10.4}}


\put(10,20){\circle{10}}

\put(10,20){\makebox(0,0)[cc]{$C$}}


\put(40,50){\circle{10}}

\put(40,50){\makebox(0,0)[cc]{$E$}}

\put(40,20){\circle{10}}

\put(40,20){\makebox(0,0)[cc]{$G$}}


\end{picture}
\end{center}
\vskip-1cm \caption{
    \label{fig2}
    The states $(A,B,C,D)$, $(A,B,E,F)$ and $(A,C,E,G)$
    of three faces of the cubic cell in Fig.~\ref{fig1}.
    The arrows labeled by 1, 2 and 3 specify the
    spanning coordinate axes for each of the faces.
    The face diagonals are depicted as thin straight lines and are mnemonically referenced
    by triples $(jk\ell)$ (with commas omitted for brevity), where $1\< j < k
    \< 3$ and $\ell = \diagdown, \diagup$, and the end sites of the face diagonals are
    ordered ``from left to right''. For example,
    the states of the face-diagonals $(1 3 \diagup)$ and $(23\diagdown)$
    are the pairs $(B,E)$ and $(A,G)$, respectively.
}
\end{figure}

Similarly to the edge-to-site entropy reduction, NNE allows each of the {\it
face entropies}
\begin{align}
\label{Hface12}
    \bH_{\face}^{(12)}
    & :=
    \bH(A,B,C,D),\\
\label{Hface13}
    \bH_{\face}^{(13)}
    & :=
    \bH(A,B,E,F),\\
\label{Hface23}
    \bH_{\face}^{(23)}
    & :=
    \bH(A,C,E,G)
\end{align}
to be reduced to the corresponding pair of \emph{face-diagonal entropies}
\begin{align}
\label{Hdiag12}
    \bH_{\diag}^{(12\diagdown)}
    & :=
    \bH(A,D),
    \qquad
    \bH_{\diag}^{(12\diagup)}
     :=
    \bH(B,C), \\
\label{Hdiag13}
    \bH_{\diag}^{(13\diagdown)}
    & :=
    \bH(A,F),
    \qquad
    \bH_{\diag}^{(13\diagup)}
     :=
    \bH(B,E), \\
\label{Hdiag23}
    \bH_{\diag}^{(23\diagdown)}
    & :=
    \bH(A,G),
    \qquad
    \bH_{\diag}^{(23\diagup)}
     :=
    \bH(C,E),
\end{align}
where the meaning of the superscripts is elucidated by Fig.~\ref{fig2}.
Here, the face entropies are defined by
\begin{equation}
\label{H_face^jk}
    \bH_{\face}^{(jk)}
    :=
    -
    \sum_{
        u,v,x,y
        \in
        \Omega
    }
    \Lambda(
        \sF_{uvxy}^{(jk)}
    ),
\end{equation}
where the degree four arrays $\sF^{(jk)} := (\sF_{uvxy}^{(jk)})_{u,v,x,y \in
\Omega}$ describe the joint PMFs of the quadruples $(A,B,C,D)$, $(A,B,E,F)$
and $(A,C,E,G)$ by
\begin{align}
\label{face_PMF_12}
    \sF_{abcd}^{(12)}
    & :=
    \bP(
        A=a,
        B=b,
        C=c,
        D=d
    ),\\
\label{face_PMF_13}
    \sF_{abef}^{(13)}
    & :=
    \bP(
        A=a,
        B=b,
        E=e,
        F=f
    ),\\
\label{face_PMF_23}
    \sF_{aceg}^{(23)}
    & :=
    \bP(
        A=a,
        C=c,
        E=e,
        G=g
    ).
\end{align}
The array $\sF^{(jk)}$ is referred to as the $(j,k)$th \emph{face
PMF}. Accordingly, the face-diagonal entropies in
(\ref{Hdiag12})--(\ref{Hdiag23}) are defined by
\begin{equation}
\label{Hdiagjkl}
    \bH_{\diag}^{(jk\ell)}
     :=
    -
    \sum_{u, y \in \Omega}
    \Lambda(\sD_{uy}^{(jk\ell)}),
\end{equation}
where the matrices $\sD^{(jk\ell)} = (\sD_{uy}^{(jk\ell)})_{u,y \in
\Omega}$ describe the corresponding \emph{face-diagonal PMFs} by
\begin{align}
\label{face_diag_PMF_12_down}
    \sD_{ad}^{(12\diagdown)}
    & :=
    \bP(
        A=a,
        D=d
    ),\\
\label{face_diag_PMF_12_up}
    \sD_{bc}^{(12\diagup)}
    & :=
    \bP(
        B=b,
        C=c
    ),\\ 
\label{face_diag_PMF_13_down}
    \sD_{af}^{(13\diagdown)}
    & :=
    \bP(
        A=a,
        F=f
    ),\\
\label{face_diag_PMF_13_up}
    \sD_{be}^{(13\diagup)}
    & :=
    \bP(
        B=b,
        E=e
    ),\\ 
\label{face_diag_PMF_23_down}
    \sD_{ag}^{(23\diagdown)}
    & :=
    \bP(
        A=a,
        G=g
    ),\\
\label{face_diag_PMF_23_up}
    \sD_{ce}^{(23\diagup)}
    & :=
    \bP(
        C=c,
        E=e
    ).
\end{align}
Associated with the $(j,k)$th face in Fig.~\ref{fig2} are two \emph{face diagonal
occupancy indicators} (FDOIs) $\omega^{(jk\diagdown)}$ and
$\omega^{(jk\diagup)}$. Each of them is a Boolean
random variable which indicates whether at least one of the end
sites of the corresponding face diagonal is occupied. More
precisely,
\begin{align}
\label{FDOI12_down}
    \omega^{(12\diagdown)}
    & :=
    \cI_W(A) \vee \cI_W(D),\\
\label{FDOI12_up}
    \omega^{(12\diagup)}
     & :=
    \cI_W(B) \vee \cI_W(C),\\
\label{FDOI13_down}
    \omega^{(13\diagdown)}
    & :=
    \cI_W(A) \vee \cI_W(F),\\
\label{FDOI13_up}
    \omega^{(13\diagup)}
     & :=
    \cI_W(B) \vee \cI_W(E),\\
\label{FDOI23_down}
    \omega^{(23\diagdown)}
    & :=
    \cI_W(A) \vee \cI_W(G),\\
\label{FDOI23_up}
    \omega^{(23\diagup)}
     & :=
    \cI_W(C)\vee \cI_W(E),
\end{align}
where $\vee$ denotes the logical disjunction, and use is made of
the SOIs $\cI_W(A), \ldots, \cI_W(H)$ mentioned in
Section~\ref{sec:edge-to-site}; see also Fig.~\ref{fig2}.

The NNE constraint implies that the FDOIs, associated with any given face of the cubic
cell, are also mutually exclusive. That is,
\begin{equation}
\label{FDOI_exclusion}
        \omega^{(jk\diagdown)}
        \nand
        \omega^{(jk\diagup)}
        =
        1
\end{equation}
for all $1\< j < k\< 3$. Moreover, this property can be used as an equivalent reformulation of the NNE constraint.
Indeed, the equivalence between the FDOI exclusion
(\ref{FDOI_exclusion}) and NNE follows from the identity $
    (\alpha \vee \delta)
    \nand
    (\beta \vee \gamma)
    =
    (\alpha \nand \beta)
    \wedge
    (\alpha \nand \gamma)
    \wedge
    (\beta \nand \delta)
    \wedge
    (\gamma\nand\delta)
$ for Boolean variables $\alpha$, $\beta$, $\gamma$, $\delta$.

Therefore, by considering admissible \emph{face configurations} which satisfy
NNE, it can be shown that the face PMFs in
(\ref{face_PMF_12})--(\ref{face_PMF_23}) are
expressed in terms of the face-diagonal PMFs in
(\ref{face_diag_PMF_12_down})--(\ref{face_diag_PMF_23_up}) as
\begin{align}
\label{F0000^jk}
    \sF_{0000}^{(jk)}
    & =
    \sD_{00}^{(jk\diagdown)}
    +
    \sD_{00}^{(jk\diagup)} - 1,\\ 
\label{Fu000^jk}
    \sF_{u000}^{(jk)}
    & =
    \sD_{u0}^{(jk\diagdown)},\\
\label{F000y^jk}
    \sF_{000y}^{(jk)}
    & =
    \sD_{0y}^{(jk\diagdown)},\\
\label{Fu00y^jk}
    \sF_{u00y}^{(jk)}
    & =
    \sD_{uy}^{(jk\diagdown)},\\ 
\label{F0v00^jk}
    \sF_{0v00}^{(jk)}
    & =
    \sD_{v0}^{(jk\diagup)},\\
\label{F00x0^jk}
    \sF_{00x0}^{(jk)}
    & =
    \sD_{0x}^{(jk\diagup)},\\
\label{F0vx0^jk}
    \sF_{0vx0}^{(jk)}
    & =
    \sD_{vx}^{(jk\diagup)},
    \qquad
    u, v, x, y \in W.
\end{align}
Hence, in order to establish the above mentioned reduction of the face entropies to the
face-diagonal entropies,
 it
now remains to substitute (\ref{F0000^jk})--(\ref{F0vx0^jk})
into the right-hand side of (\ref{H_face^jk}), so that
\begin{align}
\nonumber
    \bH_{\face}^{(jk)}
     = &
    \bH_{\diag}^{(jk\diagdown)}
    +
    \Lambda(\sD_{00}^{(jk\diagdown)})\\
\nonumber
    &+
    \bH_{\diag}^{(jk\diagup)}
    +
    \Lambda(\sD_{00}^{(jk\diagup)})\\
\nonumber
    &-
    \Lambda(\sF_{0000}^{(jk)})\\
\label{Hfacediag}     = &
    \bH_{\diag}^{(jk\diagdown)}
    +
    \bH_{\diag}^{(jk\diagup)}
    -
    \Phi^{(jk)}.
\end{align}
Here,
use is also made of (\ref{Hdiagjkl})  and (\ref{Phi}) together with
the Shannon information between the FDOIs associated with the $(j,k)$th face of the cubic cell:
\begin{align}
\nonumber
    \Phi^{(jk)}
     :=&
    \Phi(
        \sD_{00}^{(jk\diagdown)},
        \sD_{00}^{(jk\diagup)}
    )\\
\label{Phijk}
    =&
    \bI(
        \omega^{(jk\diagdown)};
        \omega^{(jk\diagup)}
    ).
\end{align}
The \emph{face-to-diagonal entropy reduction}, described by (\ref{Hfacediag}),
implies that the quantity $\Phi^{(jk)}$ is the NNE-induced Shannon information not
only between the FDOIs $\omega^{(jk\diagdown)}$ and $\omega^{(jk\diagup)}$ but
also between the states of diagonals of the $(j,k)$th face in Fig.~\ref{fig2}:
\begin{align}
\label{Phi12}
    \Phi^{(12)}
    & =
    \bI(A,D; B,C),\\
\label{Phi13}
    \Phi^{(13)}
    & =
    \bI(A,F; B,E),\\
\label{Phi23}
    \Phi^{(23)}
    & =
    \bI(A,G; C,E).
\end{align}

\section{Marginalization}
    \label{sec:face_PMF_equations}

The homogeneity of the equilibrium random field $\xi$ implies that, for every
$1\< j < k \< 3$,  the $(j,k)$th  face PMF $\sF^{(jk)}$, given by (\ref{face_PMF_12})--(\ref{face_PMF_23}),
marginalizes to the common edge PMF $\sE$ in (\ref{edge_PMF}) as
\begin{align}
\label{uv++}
    \sF_{uv\op\op}^{(jk)}
    & =
    \sE_{uv},\\
\label{++xy}
    \sF_{\op\op xy}^{(jk)}
    & =
    \sE_{xy},\\
\label{u+x+}
    \sF_{u\op x\op}^{(jk)}
    & =
    \sE_{ux},\\
\label{+v+y}
    \sF_{\op v\op y}^{(jk)}
    & =
    \sE_{vy},
    \qquad
    u,v,x,y \in \Omega.
\end{align}
Here, for the sake of brevity, the subscript $\op$ denotes the independent summation over
the corresponding dimension of an array. For example,
$\sF_{uv\op\op}^{(jk)}:= \sum_{x, y \in \Omega}\sF_{uvxy}^{(jk)}$.
The relations (\ref{uv++}) and (\ref{++xy}) imply that
$\sF_{uv\op\op}^{(jk)}=\sF_{\op\op uv}^{(jk)}$, which reflects the local
invariance of the face PMF under translations along coordinate axis
$k$. Similarly, the equality $\sF_{u\op x\op}^{(jk)}=\sF_{\op u \op x}^{(jk)}$, which follows from (\ref{u+x+}) and (\ref{+v+y}),  is related to the translational invariance along axis $j$.

By substituting (\ref{F0000^jk})--(\ref{F0vx0^jk}) for the face
PMF and (\ref{edge_PMF}) for the edge PMF, it follows that, in the NNE setting,
(\ref{uv++})--(\ref{+v+y}) are equivalent to the
marginalization of the face-diagonal PMFs to the site PMF
\begin{equation}
\label{marginal_down}
    \sD_{u\op}^{(jk\ell)}
      =
    \sD_{\op u}^{(jk\ell)}
    =
    \sS_u,
    \qquad
    u \in \Omega,
\end{equation}
for all $1\< j < k \< 3$ and $\ell =
\diagdown, \diagup$, in combination with the inequality
\begin{equation}
\label{linking}
    \sD_{00}^{(jk\diagdown)}
    +
    \sD_{00}^{(jk\diagup)}
    \> 1.
\end{equation}
This inequality originates from (\ref{F0000^jk}) and couples
the two (otherwise independent) sets of linear equations in
(\ref{marginal_down}) for different values of $\ell =
\diagdown, \diagup$, according to which the row and column sums
of the matrices $\sD^{(jk\diagdown)}$ and $\sD^{(jk\diagup)}$
reproduce the vector $\sS$.

\section{Face-diagonal admissibility}
    \label{sec:admissibility}

Some of the face-diagonal configurations may be prohibited as a result of structural restrictions, additional to NNE, such as spatial compatibility of coordination entities at the end sites
of the face diagonal (for example, if the entities overlap).
Irrespective of the nature of these additional geometric constraints, they
are described by six Boolean matrices of \emph{face-diagonal admissibility}
\begin{equation}
\label{sA}
        \sA^{(jk\ell)}
        :=
        (\sA_{uv}^{(jk\ell)})_{u,v\in \Omega},
        \qquad
        1\< j<k\< 3,\
        \ell
        =
        \diagdown, \diagup.
\end{equation}
Here, $\sA_{uv}^{(jk\ell)}$ indicates whether the configuration
$(u,v)$ is allowed for the $(j,k,\ell)$th face-diagonal; see
Fig.~\ref{fig2}. In particular, since the vacancy-vacancy configurations are admissible, then
$$
    \sA_{00}^{(jk\ell)} = 1.
$$
Each of the face-diagonal PMFs $\sD^{(jk\ell)}$ is \emph{dominated}
by the corresponding admissibility  matrix $\sA^{(jk\ell)}$ in (\ref{sA}) in the
sense that their entries satisfy the implication
\begin{equation}
\label{dominance}
    \sA_{uv}^{(jk\ell)} = 0
    \
    \Longrightarrow
    \
    \sD_{uv}^{(jk\ell)} = 0.
\end{equation}
In probability theoretic terms,  this means the absolute
continuity \cite{Shiryaev} of $\sD^{(jk\ell)}$ with respect to
$\sA^{(jk\ell)}$ which is also considered as a measure on the set $\Omega^2$. Equivalently, $\sD^{(jk\ell)}$ is concentrated on
the set
\begin{equation}
\label{admissible_diag}
    \Omega^{(jk\ell)}
    :=
    \big\{
        (u,v)\in \Omega^2:\
        \sA_{uv}^{(jk\ell)} = 1
    \big\}
\end{equation}
which consists of admissible configurations for the $(j,k,\ell)$th face
diagonal.

\section{Face-to-diagonal reduction of
    Kramers-Wannier approximation}
    \label{sec:face-to-diagonal_KW}

The entropy of the equilibrium random field $\xi$ per site of
the simple cubic lattice is defined by
\begin{equation}
\label{entropy_density}
    \bh
    :=
    \lim_{N \to+\infty}
    \frac
    {\bH(\xi_{Q_N})}
    {N^3},
\end{equation}
where $\xi_{Q_N} := (\xi_z)_{z\in Q_N}$ is the restriction of $\xi$ to the discrete cube $ Q_N$ given by (\ref{QN}).
Following the terminology
of~\cite{Pelizzola}, we will refer to $\bh$ as the \emph{entropy density}  in
order to emphasize that the configurational entropy in
(\ref{entropy_density}) is averaged per lattice site, similarly to (\ref{Slim}). An upper bound for
$\bh$ is provided by the site entropy $\bH_{\site}$ which corresponds to the
Bragg-Williams approximation~\cite{Huang}.

Recalling (\ref{Hedge}), (\ref{Hsite}) and
(\ref{Hface12})--(\ref{Hface23}), the Cluster Variation Method (CVM) with faces
of cubic cells of $\mZ^3$ as basic clusters and their edges and sites as
subclusters, known as the Kramers-Wannier approximation~\cite{Huang}, employs
the entropy density estimate
\begin{equation}
\label{hface0}
    \wh{\bh}_{\face}
    :=
    7\bH_{\site}
    -
    9
    \bH_{\edge}
    +
    \sum_{1\< j < k \< 3}
    \bH_{\face}^{(jk)}.
\end{equation}
The latter takes into account the structure of the simple cubic
lattice and the equilibrium statistical correlations in the particle
system within faces of cubic cells through the appropriate
Kikuchi-Barker coefficients \cite{Kikuchi_2002} weighting the
subcluster entropies. In contrast to $\bH_{\site}$, the quantity
$\wh{\bh}_{\face}$ is not necessarily an upper bound for the
entropy density in (\ref{entropy_density}) nor is prevented from taking negative values.
However, its advantage, even in comparison with the
Bethe-Peierls entropy density estimate, is that
$\wh{\bh}_{\face}$ ``captures'' more distant spatial
correlations.

Due to the edge-to-site and face-to-diagonal entropy reductions in
(\ref{Hedgesite}) and (\ref{Hfacediag}), which hold  in the NNE setting, (\ref{hface0})
takes the form
\begin{align}
\nonumber
    \wh{\bh}_{\face}
     = &
    9\varphi(\sS_0)
    -11 \bH_{\site}\\
\label{hface1}
    & +
    \sum_{1\< j < k \< 3}
    \left(
         -
         \Phi^{(jk)}
         +
        \sum_{\ell = \diagdown, \diagup}
         \bH_{\diag}^{(jk\ell)}
    \right),
\end{align}
which
describes a \emph{face-to-diagonal reduction} of the Kramers-Wannier
entropy density approximation in the NNE framework, where use is also made of (\ref{Phijk})--(\ref{Phi23}).

\section{Internal energy}
    \label{sec:internal_energy}

The internal energy of the particle system is modelled by the sum of
interaction energies over face diagonals of cubic cells. The
energetics is parameterized by the six energy matrices
\begin{equation}
\label{sU}
    \sU^{(jk\ell)}
    :=
    (\sU_{uv}^{(jk\ell)})_{u,v\in \Omega},
    \qquad
        1\< j<k\< 3,\
        \ell
        =
        \diagdown, \diagup.
\end{equation}
Here, $\sU_{uv}^{(jk\ell)}$ denotes the interaction energy which is ascribed to the
configuration $(u,v)$ for the $(j,k,\ell)$th face diagonal, with
\begin{equation}
\label{U00}
    \sU_{00}^{(jk\ell)}
    :=
    0,
\end{equation}
so that \emph{vacancy-vacancy} configurations are endowed with zero
energy. Hence, the average internal energy of the particle system
per lattice site is computed as
\begin{equation}
\label{bU}
    \bU
    :=
    \sum_{1\< j < k \< 3,\ \ell = \diagdown, \diagup}
        \bra
             \sU^{(jk\ell)},
             \sD^{(jk\ell)}
        \ket,
\end{equation}
in terms of the six-tuple
\begin{equation}
\label{sD}
    \sD
    :=
    \left\{
        \sD^{(jk\ell)}:\
        1\< j<k\< 3,\
        \ell
        =
        \diagdown, \diagup
    \right\}
\end{equation}
of the face-diagonal PMFs from
(\ref{face_diag_PMF_12_down})--(\ref{face_diag_PMF_23_up}), with
$\bra\cdot, \cdot\ket$ denoting the Frobenius inner product of matrices:
\begin{eqnarray*}
        \bra
             \sU^{(jk\ell)},
             \sD^{(jk\ell)}
        \ket
        & = &
        \sum_{u,y \in \Omega}
        \sU_{uy}^{(jk\ell)}
        \sD_{uy}^{(jk\ell)}.
\end{eqnarray*}
In particular, if the energy matrices in (\ref{sU}) are representable as
\begin{equation}
\label{site_energy}
    \sU_{uy}^{(jk\ell)}
    =
    \frac
    {\sU_u^{\site} +\sU_y^{\site}}
    {12},
\end{equation}
where $\sU := (\sU_a)_{a\in \Omega}$ is a single site energy function
satisfying $\sU_0 = 0$, then (\ref{bU}) reduces to
\begin{equation}
\label{singlet}
    \bU
    =
    \bra
        \sU,
        \sS
    \ket
    =
    \sum_{w\in W}
    \sU_w^{\site} \sS_w.
\end{equation}
The latter internal energy density corresponds to the model which is considered in
\cite{Vladimirov}. Note that the denominator on the right-hand side of
(\ref{site_energy}) takes into account the property that every site of the
simple cubic lattice is shared by twelve face diagonals.

\section{Gibbs energy estimate}
\label{sec:Gibbs_energy_estimate}

In the framework of the NNE-induced face-to-diagonal reduction of
the Kramers-Wannier entropy density estimate described by
(\ref{hface1}) and with the internal energy density given by
(\ref{bU}), the Gibbs free energy of the particle system per
\emph{occupied} site of the simple cubic lattice or, equivalently,
per central atom, is approximated by
\begin{equation}
\label{gface}
    \wh{\bg}_{\face}
    :=
    \min_{\sD, \sS}
    \frac
    {\bU + PV_1 - k_{\rB} T \wh{\bh}_{\face}}
    {1-\sS_0},
\end{equation}
where the denominator originates from (\ref{N1}). Here,  the minimum is
taken over the six-tuple $\sD$ of face-diagonal PMFs in
(\ref{sD}), satisfying the marginalization and admissibility constraints
(\ref{marginal_down}), (\ref{linking}) and (\ref{dominance}),
and over the site PMF $\sS$ subject to the balance equations
(\ref{cond_balance}) for given values of the pressure $P$, absolute temperature
$T$ and centralized mole fractions $y_2, \ldots, y_n$ in
(\ref{Ni/N1}).

Furthermore, $V_1$ in (\ref{gface}) denotes the physical volume of the cubic cell of the
carrier lattice and is assumed to be constant, so that $\bU + PV_1$ is the
\emph{enthalpy density} of the particle system per lattice site. The
Boltzmann constant $k_{\rB} = 1.381\x 10^{-23}$J/K converts the
information theoretic $ \wh{\bh}_{\face}$ to the thermodynamic
entropy density estimate $k_{\rB} \wh{\bh}_{\face}$, with
only the \emph{configurational} part of the entropy being taken into
account.

The solution of the constrained minimization problem
(\ref{gface}) will be described under an isotropy assumption of
the next section, which further decreases  the dimensionality of the
problem by an order of magnitude.

\section{Isotropy equivalence classes}
    \label{sec:isotropy}

In addition to the NNE constraint, suppose the six-tuple of energy matrices in
(\ref{sU})  is isotropic, that is,  invariant with respect to the full
octahedral symmetry group ${\sf O_h}$ of the cube \cite{Cotton}
consisting of 48 isometries of the simple cubic lattice generated by
mirror reflections and discrete rotations. In this case, the energy
matrices $\sU^{(jk\ell)}$ are obtained by permuting the entries
of $\sU^{(12\diagdown)}$ and are parameterized by the interaction
energies $\cU_0, \ldots, \cU_d$ which are ascribed to elements of the isotropy
equivalence classes
\begin{equation}
\label{face_diag_class}
    \Omega_0^{\diag},
    \ldots,
    \Omega_d^{\diag}
\end{equation}
into which the set $\Omega^{(12\diagdown)}$ of admissible configurations for
the $(1, 2, \diagdown)$th face diagonal  in (\ref{admissible_diag}) is
split by the action of the isometry group ${\sf O_h}$. More precisely,
\begin{equation}
\label{cU}
    \sU_{uv}^{(12\diagdown)}
    =
    \cU_i,
    \qquad
    (u,v) \in \Omega_i^{\diag},
    \quad
    i = 0, \ldots, d,
\end{equation}
where
\begin{equation}
\label{diag00}
    \Omega_0^{\diag} := \{(0,0)\}
\end{equation}
is a singleton consisting of the vacancy-vacancy configuration, so that
$$
    \cU_0 = 0
$$
in accordance with (\ref{U00}). The equivalence classes
$\Omega_1^{\diag}$, $\ldots,$ $\Omega_d^{\diag}$ are labelled in such a
way that the first $d_1$ of them satisfy the inclusion
\begin{equation}
\label{W00W}
    \bigsqcup_{i = 1}^{d_1}
    \Omega_i^{\diag}
    \subset
    \left(
        W\x \{0\}
    \right)
    \bigsqcup
    \left(
        \{0\} \x W
    \right)
\end{equation}
(that is, they consist of face-diagonal configurations with precisely one occupied site),
whereas the remaining $d_2 := d-d_1$ classes $\Omega_{d_1 +
1}^{\diag}, \ldots, \Omega_d^{\diag}$ satisfy
\begin{equation}
\label{WW}
    \bigsqcup_{i = d_1 + 1}^{d}
    \Omega_i^{\diag}
    \subset
    W^2
\end{equation}
and are formed from face-diagonal configurations with both sites occupied.

Assuming the absence of symmetry breaking, the isotropy of
the energy matrices is inherited by the finite-dimensional
probability distributions of the equilibrium random field $\xi$,
including its site and face-diagonal PMFs. Under the isotropy
assumption, the face-di\-a\-go\-nal PMFs are permutations of
entries of $\sD^{(12\diagdown)}$ and are completely specified by the
probabilities $\cD_0, \ldots, \cD_d$ of representatives of the {\it
face-diagonal classes} in (\ref{face_diag_class}) as
\begin{equation}
\label{cD}
    \sD_{uv}^{(12\diagdown)}
    =
    \cD_i,
    \qquad
    (u,v) \in \Omega_i^{\diag},
    \quad
    i = 0, \ldots, d.
\end{equation}
In particular,
\begin{equation}
\label{iso_D00}
    \sD_{00}^{(jk\ell)}
    =
    \cD_0
    \>
    \frac{1}{2}
\end{equation}
is the common probability of the vacancy-vacancy face-diagonal
configurations, with the inequality being the isotropic version of
(\ref{linking}).

By a similar reasoning, under the isotropy assumption, the site configurations
are equiprobable within each of the equivalence classes
\begin{equation}
\label{site_classes}
    \Omega_0^{\site},
    \ldots,
    \Omega_s^{\site}
\end{equation}
into which the site configurational space $\Omega$ in (\ref{Omega}) is
partitioned by the action of the isometry group ${\sf O_h}$. More precisely,
the site PMF $\sS$ takes the form
\begin{equation}
\label{cS}
    \sS_u
    =
    \cS_i,
    \qquad
    u \in \Omega_i^{\site},
    \quad
    i = 0, \ldots, s,
\end{equation}
where $\cS_i$ is its common value at elements of the $i$th \emph{site
class} $\Omega_i^{\site}$. Here, $\Omega_0^{\site} = \{0\}$ consists
of the vacancy, so that $\cS_0 =\sS_0$, while the remaining
site classes partition the set of occupied site configurations:
$$
    \bigsqcup_{i=1}^{s}
    \Omega_i^{\site}
    =
    W.
$$
Therefore, in the isotropic case being considered, the internal
energy density (\ref{bU}) and the entropy density estimate (\ref{hface1}) take the form
\begin{align}
\nonumber
    \bU
     = &
    6
    \bra
        \sU^{(12\diagdown)},
        \sD^{(12\diagdown)}
    \ket\\
\label{iso_U}
      = &
    6
    \sum_{i = 1}^{d}
    \delta_i
    \cU_i
    \cD_i,\\
\nonumber
    \wh{\bh}_{\face}
     = &
    9\varphi(\cS_0) - 11 \bH_{\site}\\
\label{iso_hface}
     & -
    3\varphi(\cD_0)
    +
    6\bH_{\diag}.
\end{align}
Here, use has been made of (\ref{varphi}), (\ref{Phi}), (\ref{Phijk}), (\ref{cU}) and
(\ref{iso_D00}). Furthermore,
\begin{equation}
\label{iso_Hdiag}
    \bH_{\diag}
     :=
    \bH_{\diag}^{(jk\ell)}
     =
    -
    \sum_{i=0}^{d}
    \delta_i
    \Lambda(\cD_i)
\end{equation}
is the common value of the face-diagonal entropies in (\ref{Hdiagjkl}), and
\begin{equation}
\label{iso_Hsite}
    \bH_{\site}
    =
    -
    \sum_{i=0}^{s}
    \sigma_i
    \Lambda(\cS_i),
\end{equation}
where
\begin{equation}
\label{sigma_delta}
    \sigma_i
    :=
    \# \Omega_i^{\site},
    \qquad
    \delta_j
    :=
    \# \Omega_j^{\diag}
\end{equation}
denote the cardinalities of the site and face-diagonal equivalence classes in (\ref{site_classes}) and (\ref{face_diag_class}), with $\sigma_0 =
\delta_0 = 1$.

\section{Conditioned representation}
\label{sec:conditioned}

In what follows, we will employ the decomposition \cite[p.~44]{Cover}   of the joint entropy for discrete random variables $\eta$ and $\zeta := f(\eta)$, the second of which  is a deterministic function of $\eta$:
\begin{align}
\nonumber
    \bH(\eta)
    & =
    \bH(\eta,\zeta)\\
\nonumber
     & =
     \bH(\zeta) + \bH(\eta \mid \zeta)\\
\label{HHH}
     & =
     \bH(\zeta)
     +
     \sum_{z}
     \bP(\zeta = z) \bH(\eta \mid \zeta=z),
\end{align}
where $\bH(\cdot \mid \cdot )$ denotes the conditional entropy, and the sum is taken over the range of $\zeta$. Similarly to the lines of reasoning for Eq.~(25) in
\cite{Vladimirov}, the site
entropy in (\ref{iso_Hsite}) can be represented as
\begin{align}
\label{HsiteSOI}
    \bH_{\site}
     =
    \bH_{\rm SOI}
    +
    (1-\cS_0)
    \bH(A \mid A\ne 0),
\end{align}
which is obtained by applying (\ref{HHH}) to the random variable $\eta := A$ and the SOI $\zeta := \cI_W(A)$ (see Section~\ref{sec:edge-to-site}) and using the relation
$$
    \bH(A\mid \cI_W(A)=0)=\bH(A\mid A=0) = 0.
$$
In the framework of the isotropy assumption of Section~\ref{sec:isotropy},
the common entropy of the SOIs $\cI_W(A), \ldots, \cI_W(H)$ in (\ref{HsiteSOI}) is  computed as
\begin{equation}
\label{HSOI}
    \bH_{\rm SOI}
    :=
    -\Lambda(\cS_0)
    -
    \Lambda(1-\cS_0).
\end{equation}
Also, the conditional entropy of the state of a given site of the carrier lattice,
provided that the site is occupied, takes the form
\begin{equation}
\label{iso_cond_site_entropy}
    \bH(A\mid A\ne 0)
    =
    -
    \sum_{i =1}^{s}
    \sigma_i
    \Lambda (\wt{\cS}_i),
\end{equation}
 with
\begin{equation}
\label{iso_cond_site_PMF}
    \wt{\cS}_i
     :=
    \frac
    {\cS_i}
    {1-\cS_0},
    \qquad
    i = 1, \ldots, s,
\end{equation}
describing the isotropic version of the conditional site PMF in
(\ref{cond_site_PMF}) in view of (\ref{cS}) and (\ref{sigma_delta}), so that $\sum_{i=1}^s \sigma_i \wt{\cS}_i = 1$.

A similar conditioned representation for the face-diagonal entropy
in (\ref{iso_Hdiag}) is
\begin{align}
\nonumber
    \bH_{\diag}
    = &
    \bH_{\FDOI}\\
\label{HdiagFDOI}
    & +
    (1-\cD_0)
    \bH(
        A,D
        \mid
        \omega^{(12\diagdown)} = 1
    ),
\end{align}
where
\begin{align}
\nonumber
    \bH_{\FDOI}
    := &
    \bH(\omega^{(jk\ell)})\\
\label{HFDOI}
    = &
    -\Lambda(\cD_0)
    -
    \Lambda(1-\cD_0)
\end{align}
is the common entropy of the FDOIs in
(\ref{FDOI12_down})--(\ref{FDOI23_up}) under the isotropy
assumption, and
\begin{equation}
\label{iso_cond_diag_entropy}
    \bH(
        A, D
        \mid
        \omega^{(12\diagdown)} = 1
    )
    =
    -
    \sum_{i =1}^{d}
    \delta_i
    \Lambda (\wt{\cD}_i)
\end{equation}
is the conditional entropy of the state of a given face diagonal provided that
the associated FDOI is true. Here, in accordance with
(\ref{cD}), the probabilities
\begin{equation}
\label{iso_cond_diag_PMF}
    \wt{\cD}_i
     :=
    \frac
    {\cD_i}
    {1-\cD_0},
    \qquad
    i = 1, \ldots, d,
\end{equation}
comprise the conditional face-diagonal PMF in the isotropic case,  so that $\sum_{i=1}^d \delta_i \wt{\cD}_i = 1$ in view of (\ref{sigma_delta}). Furthermore, the internal energy density in
(\ref{iso_U}) is expressed in terms of $\wt{\cD}_1,
\ldots, \wt{\cD}_d$ as
\begin{equation}
\label{iso_cond_U}
    \bU
    =
    6
    (1-\cD_0)
    \sum_{i = 1}^{d}
    \delta_i
    \cU_i
    \wt{\cD}_i.
\end{equation}

\section{Balance equations}
\label{sec:iso_balance}

For any given particle species,   the number of its representatives in a
coordination entity depends only on the site class to which this entity
belongs, so that
$$
    \Gamma_{iw}
    =
    \gamma_{ij},
    \qquad
    w \in \Omega_j^{\site}.
$$
Hence, under the isotropy assumption of Section~\ref{sec:isotropy},
the balance equations (\ref{cond_balance}) take the form
\begin{equation}
\label{iso_balance}
    \frac{1}{\nu_i}
    \sum_{j=1}^{s}
    \gamma_{ij}
    \sigma_j
    \wt{\cS}_j
    =
    y_i,
    \qquad
    i = 1, \ldots, n,
\end{equation}
where use is made of the conditional site PMF described by
(\ref{iso_cond_site_PMF}). By assembling the coefficients in (\ref{iso_balance})  into an
$(n\x s)$-matrix
\begin{equation}
\label{Ups}
    \Ups
    :=
    \left(
        \frac{\gamma_{ij}\sigma_j}{\nu_i}
    \right)_{1\< i \< n,\, 1\< j \< s},
\end{equation}
this system of linear equations can be represented in vector-matrix form as
\begin{equation}
\label{iso_balance_system}
    \Ups \wt{\cS}
    =
    Y.
\end{equation}
Here, $\wt{\cS} := (\wt{\cS}_i)_{1\< i \< s}$ and $Y
:= (y_i)_{1\< i \< n}$ are column-vectors of the probabilities from
(\ref{iso_cond_site_PMF}) and the centralized mole fractions defined
by (\ref{Ni/N1}). The first row of the matrix $\Ups$ consists of the occupied site class cardinalities:
$$
    \Ups_{1\bullet}
    =
    \begin{bmatrix}
    \sigma_1 &
    \ldots &
    \sigma_s
    \end{bmatrix}.
$$

\section{Marginalization equations}
\label{sec:iso_marginalization}

\begin{figure}[htb]
    \begin{center}
        \epsfbox{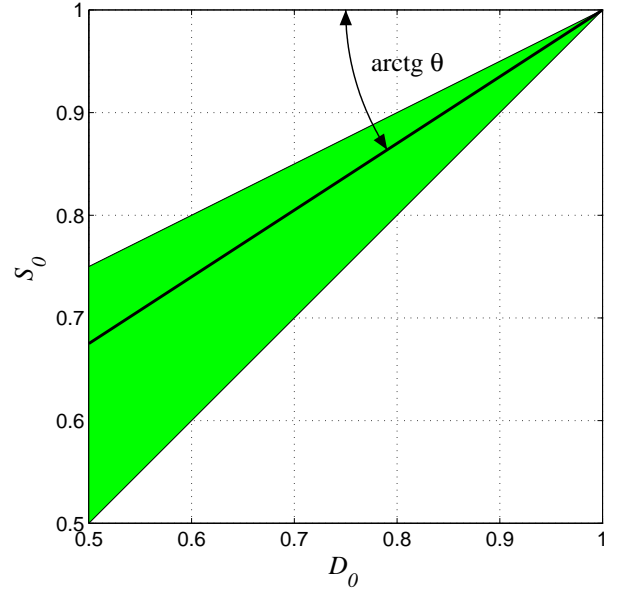}
    \end{center}
    \caption{
    \label{fig:D0_S0}
    The shaded triangle is the set of admissible values of the pair
$(\cD_0, \cS_0)$, which is  described by $\frac{1}{2}\< \cD_0 < \cS_0 \<
\frac{1}{2}(1+\cD_0)$. The bold line segment represents the set of constant
ratio $\theta$ defined by (\ref{theta}).
    }
\end{figure}

Under the isotropy assumption of Section~\ref{sec:isotropy}, the marginalization equations
(\ref{marginal_down}) can be represented in vector-matrix form
as
\begin{equation}
\label{MDLS}
    M \cD
    =
    L \cS,
\end{equation}
where $\cD := (\cD_i)_{0\< i \< d}$ and $\cS := (\cS_i)_{0\< i\<
s}$ are column-vectors with entries from (\ref{cD})
and (\ref{cS}). The matrix  $M$ is given by
\begin{equation}
\label{M}
    M
    :=
    \left[
    \begin{array}{c|c}
    \begin{array}{c}
    1\\
    \bzero_{r\x 1}
    \end{array}
    &
    \wt{M}
    \end{array}
    \right],
    \qquad
    \wt{M}
    :=
    \begin{bmatrix}
    M_0 \\
    M_1
    \end{bmatrix}
\end{equation}
and is assumed to be of full row rank. Here, $\bzero_{p\x q}$ denotes
the $(p\x q)$-matrix of zeros. Also, the matrix $L$ in (\ref{MDLS}) is block-diagonal,
\begin{equation}
\label{L}
    L
    :=
    \begin{bmatrix}
    1 & \bzero_{1\x s}\\
    \bzero_{r\x 1} & \wt{L}
    \end{bmatrix},
    \qquad
    \wt{L}
    :=
    \begin{bmatrix}
    \bone_{r_1} &  & \bzero\\
     & \ddots \\
    \bzero &  & \bone_{r_s}
    \end{bmatrix},
\end{equation}
where $\bone_p$ denotes the $p$-dimensional column-vector of
ones, and
$$
    r
    :=
    \sum_{i=1}^{s}
    r_i.
$$
The block $M_0$ of the matrix $\wt{M}$ in (\ref{M}) is a
$d$-dimensional row-vector whose first $d_1$ entries are the
half-cardinalities of the face-diagonal classes $\Omega_1^{\diag},\ldots,\Omega_{d_1}^{\diag}$ from
(\ref{W00W}) padded with $d_2$ zeros:
\begin{equation}
\label{M0}
    M_0
    :=
    \frac{1}{2}
    \begin{bmatrix}
        \delta_1 &
        \ldots &
        \delta_{d_1}
        &
        \bzero_{1\x d_2}
    \end{bmatrix}.
\end{equation}
Accordingly, the bottom block $M_1$ is an $(r\x d)$-matrix. The
structure of the matrices $M$ and $L$ allows (\ref{MDLS}) to be
represented in the form
\begin{equation}
\label{thetaMDLS}
    \wt{M}
    \wt{\cD}
    =
    \begin{bmatrix}
        1-\theta\\
        \theta\wt{L} \wt{\cS}
    \end{bmatrix}
\end{equation}
whose right-hand side depends on the probabilities $\cD_0$ and $\cS_0$ only through the auxiliary variable
\begin{equation}
\label{theta}
    \theta
    :=
    \frac{1-\cS_0}{1-\cD_0}.
\end{equation}
Here, $\wt{\cD} := (\wt{\cD}_i)_{1\< i \< d}$ is the
column-vector of probabilities from (\ref{iso_cond_diag_PMF}).
In view of (\ref{FDOI12_down}), the probabilistic meaning of
$\theta$ is clarified by
\begin{align*}
    \theta
    & =
    \frac{\bP(A\ne 0)}{\bP(\omega^{(12\diagdown)} = 1)}\\
    & =
    \frac{\bP(A\ne 0,\, \omega^{(12\diagdown)} = 1)}{\bP(\omega^{(12\diagdown)} = 1)}    \\
    & =
    \bP
    (
        A \ne 0
        \mid
        \omega^{(12\diagdown)} = 1
    ),
\end{align*}
where use is made of the property that the event $A\ne 0$ implies $\omega^{(12\diagdown)} = 1$.
In contrast to the pair $(\cD_0, \cS_0)$, whose admissible values are
depicted in Fig.~\ref{fig:D0_S0}, the quantities $\cD_0$ and
$\theta$ are functionally independent, with each of them taking values in the
interval $[1/2, 1)$.

\def\epsfsize#1#2{0.55#1}

\section{Separation of variables}
\label{sec:separation}

By dividing the numerator in (\ref{gface}) by $k_{\rB}T$, the minimization can be reduced to that of the following dimensionless function
\begin{equation}
\label{bg}
    \bg
    :=
    \frac{1}{1-\cS_0}
    \left(
        K +
        \frac{\bU}{k_{\rB} T}
        -\wh{\bh}_{\face}
    \right),
\end{equation}
where
\begin{equation}
\label{K}
    K
    :=
    \frac
    {PV_1}
    {k_{\rB} T}
\end{equation}
resembles the compressibility factor. In view of
(\ref{theta}), the probability $\cS_0$ can be expressed in terms of $\cD_0$ and
$\theta$ as
\begin{equation}
\label{SthetaD}
    \cS_0
    =
    1- (1-\cD_0) \theta.
\end{equation}
This will allow $\bg$ in (\ref{bg}) to be minimized as a function of $\cD_0$,
$\theta$, $\wt{\cD}$, $\wt{\cS}$ by employing a se\-pa\-ra\-tion-of-va\-ri\-ables
technique, similar to that in Dynamic Programming. More precisely, $\bg$ can be  split
into the sum of two functions which share $\theta$ as a common argument:
\begin{equation}
\label{ggg}
    \bg
     =
    \bg_0(\cD_0, \theta)
    +
    \bg_1(\theta,\wt{\cD}, \wt{\cS}).
\end{equation}
Since $\cD_0$
enters the balance and marginalization equations
(\ref{iso_balance_system}) and (\ref{thetaMDLS}) only through
$\theta$, then the variables $\cD_0$ and $(\wt{\cD},
\wt{\cS})$ are functionally independent for any given value
of $\theta$. Therefore, the problem of constrained minimization of $\bg$  in (\ref{ggg}) can be decomposed
as
\begin{align}
\nonumber
    \wh{\bg}
     :=&
    \min_{\cD_0, \theta, \wt{\cD}, \wt{\cS}} \bg\\
\label{hatggg}
    =&
    \min_{1/2\< \theta < 1}
    \left(
        \wh{\bg}_0(\theta)
        +
        \wh{\bg}_1(\theta)
    \right)
\end{align}
into the optimization problems
\begin{align}
\label{hatg0}
    \wh{\bg}_0(\theta)
    := &
    \min_{1/2\< \cD_0 < 1}
    \bg_0(\cD_0, \theta),\\
\nonumber
    \wh{\bg}_1(\theta)
    := &
    \min
    \{
        \bg_1(\theta,\wt{\cD}, \wt{\cS}): \\
\label{hatg1}
    &
        \wt{\cD}, \wt{\cS}\
        {\rm satisfy}\ (\ref{iso_balance_system})\ {\rm and}\ (\ref{thetaMDLS})
    \}
\end{align}
which are solved separately for any given value of  $\theta$ as a ``master'' parameter. The specific form of the functions $\bg_0$ and $\bg_1$
is as follows. A combination of
(\ref{iso_hface})--(\ref{sigma_delta}), (\ref{iso_cond_U}) and
(\ref{varphi}) with (\ref{bg}) yields
\begin{equation}
\label{g0}
    \bg_0(\cD_0, \theta)
     =
    \frac
    {K  + \psi(\cD_0,\, 1-(1-\cD_0)\theta)}
    {(1-\cD_0) \theta},
\end{equation}
where
\begin{align}
\nonumber
    \psi(\cD_0,\cS_0)
 := &
    3\varphi(\cD_0) - 6\bH_{\FDOI}\\
\nonumber
    & -
    9 \varphi(\cS_0) + 11 \bH_{\rm SOI}\\
\nonumber
     = &
    3 \Lambda(2\cD_0-1) + 6\Lambda(1-\cD_0)\\
\nonumber
    & -
    9\Lambda(2\cS_0-1) + 7\Lambda(\cS_0)\\
\label{psi}
    &  -
    11 \Lambda(1-\cS_0),
\end{align}
and $\cS_0$ is expressed in terms of $\cD_0$ and $\theta$ by
(\ref{SthetaD}). A similar reasoning leads to
\begin{align}
\nonumber
    \bg_1(\theta, \wt{\cD}, \wt{\cS})
     = &
    \frac{6}{\theta}
    \sum_{i = 1}^{d}
    \delta_i
    \left(
        \frac{\cU_i \wt{\cD}_i}{k_{\rB} T}
        +
        \Lambda(\wt{\cD}_i)
    \right)\\
\label{g1}
    & -
    11
    \sum_{j=1}^{s}
    \sigma_j
    \Lambda(\wt{\cS}_j).
\end{align}
The pressure $P$ in (\ref{K}) and the central mole fractions in
(\ref{Ni/N1}) enter the Gibbs free energy approximation in two
separate ways --- via the function $\bg_0$ in (\ref{g0}) and
through the linear constraints in (\ref{hatg1}) which specify the
function $\wh{\bg}_1$. However, being  ``blended'' by the
minimization over  $\theta$ on the right-hand side of
(\ref{hatggg}), the effects of pressure and chemical composition
are not additive.

\subsection{Minimization of $\bg_0$}\label{sec:hatg0}

\begin{figure}[htb]
    \begin{center}
    \epsfbox{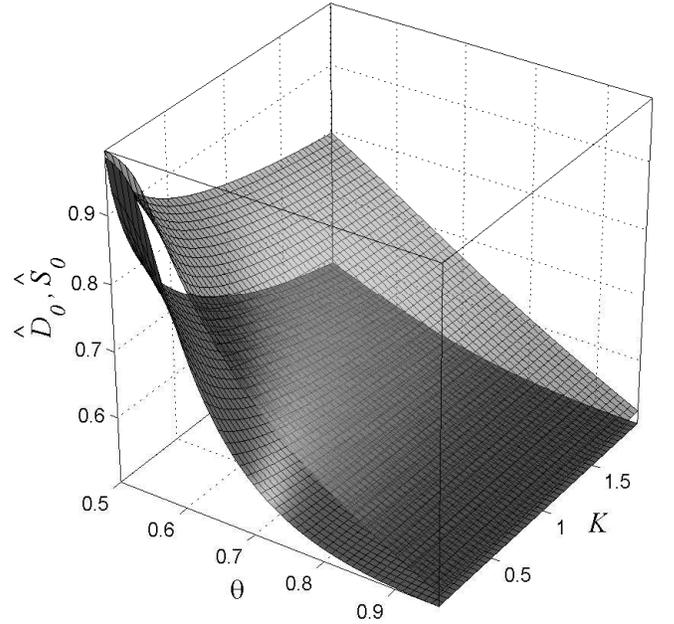}
    \end{center}
    \caption{
    \label{fig:hat_D0_S0}
    The graphs of $\wh{\cD}_0$ (lower opaque surface) and $\wh{\cS}_0$ (upper transparent surface)
    as functions of $\theta$ and $K$ defined by (\ref{decic}) and (\ref{hat_D0_S0}).
    }
\end{figure}

\begin{figure}[htb]
    \begin{center}
        \epsfbox{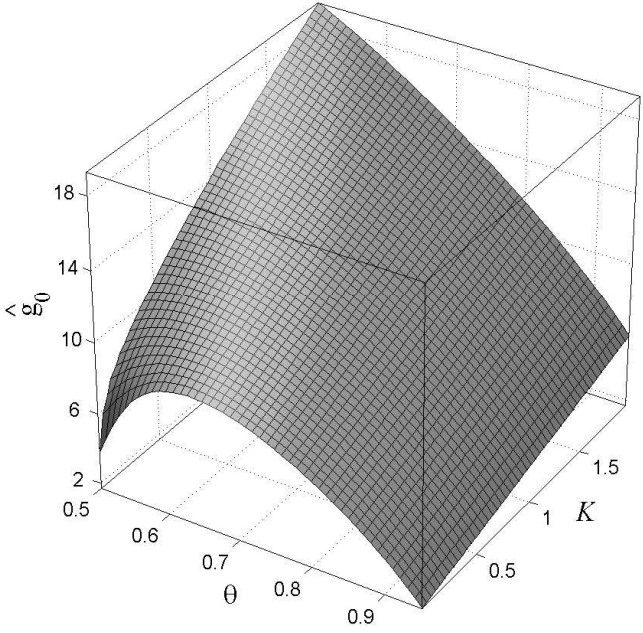}
    \end{center}
    \caption{
    \label{fig:hat_g0}
    The graph of $\wh{\bg}_0$
    defined by (\ref{hatg0}) as a function of $\theta$ and $K$.
    }
\end{figure}

A combination of (\ref{SthetaD}) with (\ref{g0}) and (\ref{psi}) yields the following partial derivative of the function $\bg_0$ with respect to the probability $\cD_0$:
\begin{align}
\nonumber
    \d_{\cD_0} \bg_0
    &=
    \frac
    {(1-\cD_0)\left(\d_{\cD_0}\psi  + \theta \d_{\cS_0}\psi\right) + K + \psi}
    {(1-\cD_0)^2\theta}\\
\nonumber
    &=
    \frac
    {(1-\cD_0)\d_{\cD_0}\psi + (1-\cS_0) \d_{\cS_0}\psi + K + \psi}
    {(1-\cD_0)^2\theta}\\
\label{dg0dD0}
    &=
    \frac
    {K + 3\ln(2\cD_0-1) - 9\ln(2\cS_0-1) + 7\ln \cS_0}
    {(1-\cD_0)^2\theta}.
\end{align}
For any given $\theta\in [1/2, 1)$, the numerator of the fraction in (\ref{dg0dD0}) tends to $-\infty$ as $\cD_0 \to \frac{1}{2}$ and approaches  the quantity $K>0$ as $\cD_0 \to
1$. Hence, by the Intermediate Value Theorem, this ensures solvability of
the equation $\d_{\cD_0} g_0 = 0$ on the interval and
achievability of the minimum in (\ref{hatg0}). Moreover, the minimum
is achieved at a unique point which is related to the
appropriate root $
    \chi := 1-\cD_0
    \in (0, 1/2]
$ of the decic equation
\begin{equation}
\label{decic}
    \left(
        \chi - \frac{1}{2}
    \right)^3
    \left(
        \chi - \frac{1}{\theta}
    \right)^7
    +
    64\theta^2\re^{-K}
    \left(
        \chi - \frac{1}{2\theta}
    \right)^9
    =
    0,
\end{equation}
whose unique solvability is established in
Appendix~\ref{sec:unique}. The corresponding values of the
probabilities $\cD_0$ and $\cS_0$ are given by
\begin{equation}
\label{hat_D0_S0}
    \wh{\cD}_0 := 1- \chi,
    \qquad
    \wh{\cS}_0 := 1 - \chi\theta
\end{equation}
and are depicted, together with $\wh{\bg}_0$,  as functions of
$\theta$ and $K$ in Figs.~\ref{fig:hat_D0_S0} and \ref{fig:hat_g0}.
\def\epsfsize#1#2{0.6#1}
\def\epsfsize#1#2{0.6#1}
For any value of the parameter $K > 0$ in (\ref{K}), the root
$\chi$ is a smooth function of $\theta \in [1/2, 1)$ and so also are
$\wh{\cD}_0$, $\wh{\cS}_0$ and $\wh{\bg}_0$. Since
$\d_{\cD_0} \bg_0$ vanishes at $\cD_0 =
\wh{\cD}_0$, then
\begin{align}
\nonumber
    \wh{\bg}_0'(\theta)
    & =
    \d_{\theta}\bg_0(\cD_0, \theta)
    \Big|_{\cD_0 = \wh{\cD}_0}\\
\label{hatg0dash}
    & =
    -
    \frac{(1-\wh{\cS}_0) \d_{\cS_0}\psi+
    K+\psi}{(1-\wh{\cD}_0)\theta^2 }.
\end{align}
From data on the molar volume of magmatic liquids in Fig.~2.3 on
p.~29 of \cite{Mysen}, a typical value of the cell volume is $V_1 \sim 10^{-29}$m$^3$, and hence, an estimate of the quantity $K$ in (\ref{K}) at $P \sim 10^5$Pa and
$T \sim 10^3$K is $K \sim 10^{-4}$. Thus,
typically being
small, $K$ becomes close to 1 for tectonic pressures of $10^8$ to
$10^9$ Pa.

\subsection{Minimization of $\bg_1$}\label{sec:Lagrange}

Although the function $\bg_1(\theta, \wt{\cD}, \wt{\cS})$ in
(\ref{g1}) is strictly convex with respect to $\wt{\cD}$, it is
strictly concave with respect to  $\wt{\cS}$. This saddle-like landscape (where there is a guarantee only for a unique
minimum over $\wt{\cD}$ at a given $\wt{\cS}$)
complicates the constrained minimization problem in
(\ref{hatg1}). The
associated Lagrange function is given by
\begin{align}
\nonumber
    \cL(\theta, \wt{\cD}, \wt{\cS}, \lambda_0, \lambda, \mu)
     = &
    \bg_1(\theta, \wt{\cD}, \wt{\cS})
    \\
\nonumber
    & -
    \lambda_0 (M_0 \wt{\cD} - (1-\theta))\\
\nonumber
    & -
    \bra
         \lambda,\,
         M_1
         \wt{\cD}
         -
         \theta\wt{L} \wt{\cS}
    \ket\\
\label{cL}
    & -
    \bra
        \mu,\,
        \Ups \wt{\cS} - Y
    \ket.
\end{align}
Here, the scalar $\lambda_0$ and the column-vectors $\lambda :=
(\lambda_j)_{1\< j \< r}$ and $\mu := (\mu_k)_{1 \< k \< n}$ are
the Lagrange multipliers associated with the marginalization and
balance equations (\ref{thetaMDLS}) and (\ref{iso_balance_system}), and
use is made of the structure of the matrix $\wt{M}$
in (\ref{M}). By substituting (\ref{g1}) into
(\ref{cL}), it follows that the condition of stationarity of the Lagrange function with respect to
$\wt{\cD}$ takes the form
\begin{align}
\nonumber
    \d_{\wt{\cD}_i}\cL
     = &
    \frac{6 \delta_i}{\theta}
    \left(
        \frac{\cU_i }{k_{\rB} T}
        +
        1 + \ln \wt{\cD}_i
    \right)\\
\label{dLdD}
    & -
    \lambda_0 (M_0)_i
    -
    \bra
        \lambda,\,
        (M_1)_{\bullet i}
    \ket
     = 0,
    \qquad
    i = 1, \ldots, d,
\end{align}
where $(M_0)_i$ denotes the $i$th entry of the row-vector $M_0$ in (\ref{M0}) and $(M_1)_{\bullet i}$ is the $i$th column of the matrix $M_1$. In a similar vein, the equations of stationarity of $\cL$ with respect to $\wt{\cS}$ are
\begin{align}
\nonumber
    \d_{\wt{\cS}_j} \cL
     = &
    -11
    \sigma_j
    \left(
        1 + \ln \wt{\cS}_j
    \right)\\
\label{dLdS}
    & +
    \theta
    \bra
        \lambda,
        \wt{L}_{\bullet j}
    \ket
    -
    \bra
        \mu,
        \Ups_{\bullet j}
    \ket
     =
    0,
    \qquad
    j = 1, \ldots, s,
\end{align}
where $\Ups_{\bullet j}$ and $\wt{L}_{\bullet j}$ denote the
$j$th columns of the matrices $\Ups$ and $\wt{L}$ in
(\ref{Ups}) and (\ref{L}), respectively. Hence,
$\wt{\cD}$ and $\wt{\cS}$ are expressed in terms of
the Lagrange multipliers $\lambda_0$, $\lambda$ and $\mu$ as
\begin{align*}
\wt{\cD}_i
     = &
    \exp
    \left(
        \frac{\theta}{6\delta_i}
        \left(
            \lambda_0 (M_0)_i + \bra \lambda, (M_1)_{\bullet i}\ket
        \right)
        -\frac{\cU_i}{k_{\rB} T}
        -1
    \right),\\
\wt{\cS}_j
     = &
    \exp
    \left(
        \frac{1}{11\sigma_j}
        \left(
            \theta
            \bra
                \lambda,
                \wt{L}_{\bullet j}
            \ket
            -
            \bra
                \mu,
                \Ups_{\bullet j}
            \ket
        \right)
        -1
    \right).
\end{align*}
Substitution of  these expressions to (\ref{thetaMDLS}) and
(\ref{iso_balance_system}) leads to a system of $1+r+n$ nonlinear
equations for the scalar variables $\lambda_0, \ldots, \lambda_r$ and
$\mu_1, \ldots, \mu_n$, which can be solved numerically, for example,  by Newton
iterations (see Appendix~\ref{sec:newton}) organized as
two nested loops. The inner loop iterates for $\lambda_0$ and
$\lambda$ in order to achieve (\ref{thetaMDLS}), while the outer loop
solves for $\mu$ to satisfy (\ref{iso_balance_system}).

Note, however, that the development of a reliable algorithm for solving the constrained optimization problem in (\ref{hatg1}) (in particular, able to avoid false extrema) is a separate problem which needs to be solved for the computer implementation of this  approach.

The minimum value in (\ref{hatg1}),  delivered by a stationary point
$(\wh{\lambda}_0, \wh{\lambda}, \wh{\mu})$, can be computed by
multiplying both sides of (\ref{dLdD}) and (\ref{dLdS}) by
$\wt{\cD}_i$ and $\wt{\cS}_j$, respectively, and
taking the sum over $i$ and $j$:
$$
    \wh{\bg}_1(\theta)
    =
    (1-\theta)
    \wh{\lambda}_0  +
    \bra
        Y,
        \wh{\mu}
    \ket + 11 - \frac{6}{\theta},
$$
where use is also made of (\ref{g1}) and the previously mentioned identities
$
    \sum_{i = 1}^{d}
    \delta_i \wt{\cD}_i
    =
    \sum_{j = 1}^{s}
    \sigma_j \wt{\cS}_j
    =1
$.

\subsection{Minimization over the master parameter}
\label{sec:blending}

If, in combination with the stationarity condition of the previous
section, the minimum on the right-hand side of (\ref{hatggg}) is
achieved at an interior value of the master parameter $\theta$, then
\begin{align}
\nonumber
    \wh{\bg}_0'(\theta)
    +
    \d_{\theta}\cL
     = &
    \wh{\bg}_0'(\theta)
    +
    \d_{\theta}\bg_1\\
\label{blending}
    & -
    \wh{\lambda}_0
    +
    \bra
         \wh{\lambda},\,
         \wt{L} \wh{\wt{\cS}}
    \ket
    =
    0.
\end{align}
Here, $\wh{\bg}_0'$ is given by (\ref{hatg0dash}), and the
partial derivatives of the Lagrange function $\cL$ from
(\ref{cL}) and those of the function $\bg_1$ from (\ref{g1}) with respect to
$\theta$ are evaluated at the solution
$(\wh{\wt{\cD}},\,
    \wh{\wt{\cS}})$ of the constrained minimization
    problem of the previous section and the associated Lagrange
    multipliers $(\wh{\lambda}_0, \wh{\lambda})$.
    Therefore, the above mentioned two-loop scheme can be equipped with
    an outermost loop which solves for the master parameter $\theta$ to satisfy
    (\ref{blending}).

\section{An application to modelling of ${\rm SiO_2-M_2O}$}
    \label{sec:binary_silicate}

The isotropic version of the face-to-diagonal reduction of the
Kramers-Wannier approximation is applicable to the statistical
mechanical modeling of the ${\rm SiO_2}$--${\rm M_2O}$ silicate melt
in the framework of the above mentioned NNE cubic lattice structural
model~\cite{Vladimirov}, where M is a univalent metal. The data
below provide an insight into the combinatorial and numerical aspects
of this application.

\begin{figure}[htb]
    \begin{center}
        \def\epsfsize#1#2{0.65#1}
        \epsfbox{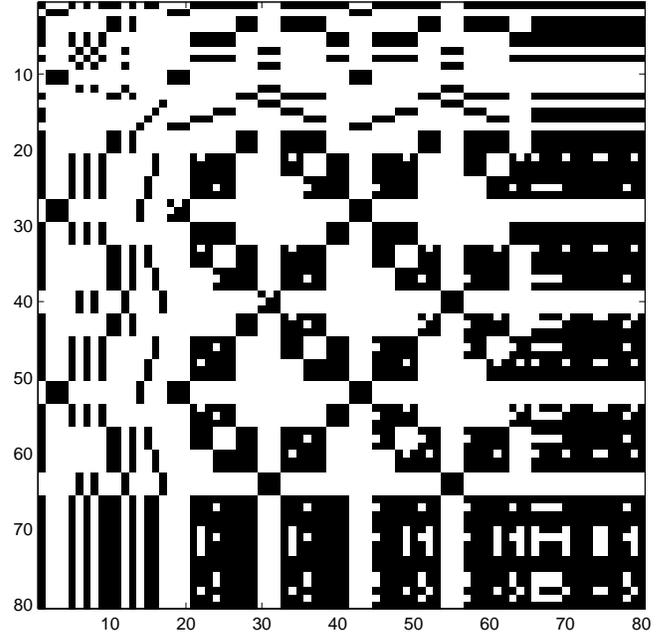}
    \end{center}
    \caption{
        \label{fig:face_diag_adj}
        The sparsity pattern of the face-diagonal
        admissibility matrix $\sA^{(12\diagdown)}$ for the ${\rm SiO_2}$--${\rm
        M_2O}$ silicate model, with its 2711 nonzero entries
        shown in black.
    }
\end{figure}

Including the oxygen vacancy, the set $\Omega$ consists of 80 site
configurations, which are split into seven isotropy equivalence
classes described in Table~\ref{tab1}.
\begin{table}
\caption{
    \label{tab1}
    The site classes $\Omega_0^{\site}, \ldots, \Omega_6^{\site}$ for the    NNE-constrained simple cubic lattice model of the ${\rm SiO_2}$--${\rm
    M_2O}$ liquid silicate, with bond angles rounded to whole degrees. Also shown are the cardinalities of the classes and the O, Si
    and M contents in their representatives.
    }
    \begin{ruledtabular}
        \begin{tabular}{|c|l|c|c|c|c|}
        $k$   & $\Omega_k^{\site}$ & $\sigma_k$ & $\gamma_{1k}$ & $\gamma_{2k}$ & $\gamma_{3k}$ \\
        \hline
    0   & oxygen vacancy   &  1        &   0   &   0  &   0     \\
    1   & Si--O--Si $\measuredangle\, 109^{\circ}$ &      12  &   1  &   2   &   0  \\
    2   & Si--O--Si $\measuredangle\, 180^{\circ}$ &   4   &   1  &   2   &   0  \\
    3   & Si--O--M $\measuredangle\, 55^{\circ} $ &   24  &   1  &   1   &   1  \\
    4   & Si--O--M $\measuredangle\, 125^{\circ}$ &   24  &   1  &   1   &   1  \\
    5   & M--O--M $\measuredangle\,  90^{\circ} $ &   12  &   1  &   0   &   2  \\
    6   & M--O--M $\measuredangle\, 180^{\circ}$ &   3   &   1  &   0   &   2  
    \end{tabular}
\end{ruledtabular}
\end{table}
By a computer-aided analysis of configurational spaces on larger
clusters (carried out using  MATLAB), each of the sets
$\Omega^{(jk\ell)}$ is split into $384$ isotropy equivalence
classes and consists of 2711 admissible face-diagonal configurations,
which is significantly less than $(\#\Omega)^2 = 6400$ due to the
geometric constraints described in Section~IV of
\cite{Vladimirov}; see also Fig.~\ref{fig:face_diag_adj}.

\begin{figure}[htb]
    \begin{center}
        \def\epsfsize#1#2{0.65#1}
        \epsfbox{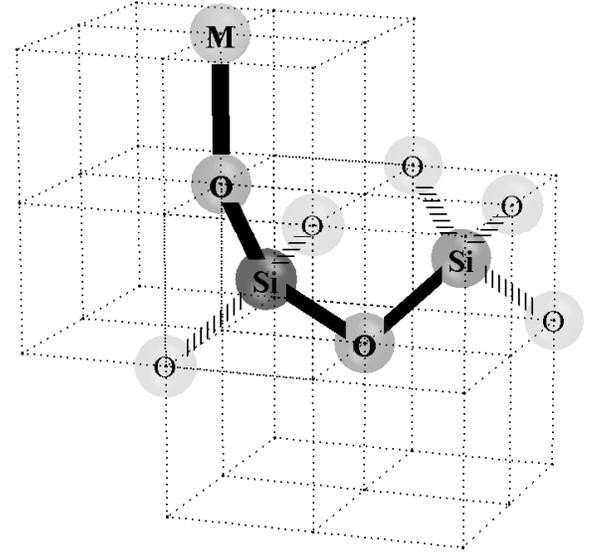}
    \end{center}
    \caption{
        \label{fig:Si_O_Si_O_M}
        A  face-diagonal configuration M--O--Si--O--Si with
        overlapping coordination entities Si--O--M and Si--O--Si which share
        a Si atom. Pale shadings represent the other parts of the silica
        tetrahedra.
    }
\end{figure}

\begin{figure}[htb]
    \begin{center}
        \def\epsfsize#1#2{0.65#1}
        \epsfbox{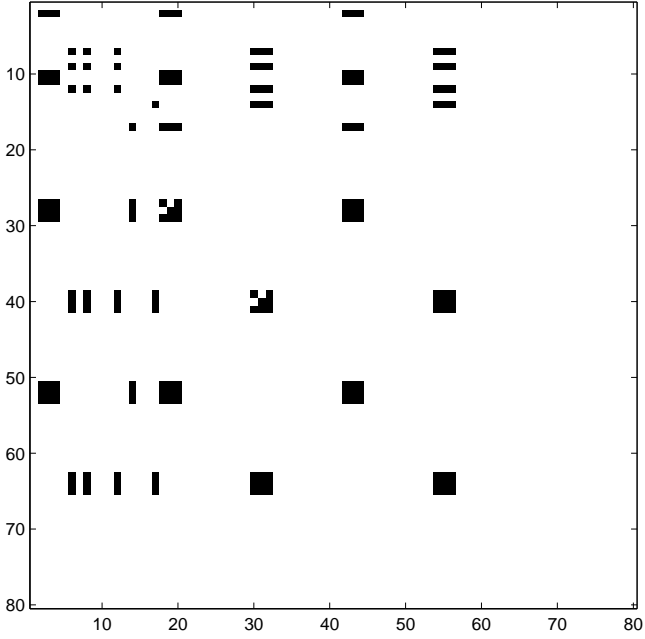}
    \end{center}
    \caption{
        \label{fig:overlap_sparsity}
        The black areas represent the 184 face-diagonal
        configurations which describe overlapping coordination
        entities associated with the end sites of the
        $(1,2,\diagdown)$th face diagonal.
    }
\end{figure}

Furthermore, 118 admissible \emph{oxygen-vacancy} face-diagonal
configurations, which belong to the union of the sets on the right-hand side of (\ref{W00W}),
are split into $d_1 = 20$
isotropy equivalence classes $\Omega_1^{\diag}, \ldots,
\Omega_{20}^{\diag}$.
The remaining 2592 \emph{oxygen-oxygen} face-diagonal configurations
belonging to $W^2$ are split into $d_2=363$ equivalence classes
$\Omega_{21}^{\diag}, \ldots, \Omega_{383}^{\diag}$.
 Among them, 184 face-diagonal configurations, partitioned into 31 equivalence
class, are \emph{overlapping} coordination entities which share a
common Si atom and, thus, can be written as X--O--Si--O--Y, where X
and Y stand for Si or M cations; see Figs.~\ref{fig:Si_O_Si_O_M} and
\ref{fig:overlap_sparsity}.

\subsection{Energy parameters}
    \label{sec:energy_model}

The energetics of the ${\rm SiO_2}$--${\rm M_2O}$ silicate model is
specified by $d = 383$ interaction energies $\cU_1, \ldots, \cU_d$
associated with representatives of the appropriate face-diagonal
classes. This is a large number of parameters.

A substantially more economical parameterization is provided by the
single site energy model described by (\ref{site_energy}) whose
isotropic version is given by
\begin{equation}
\label{iso_site_energy}
    \cU_i
    =
    \frac
    {\cU_j^{\site} + \cU_k^{\site}}
    {12},
    \qquad
    (u,y) \in \Omega_i^{\diag},
    u \in \Omega_j^{\site},\
    y \in \Omega_k^{\site},
\end{equation}
where $s = 6$ energies $\cU_1^{\site}, \ldots, \cU_s^{\site}$ are
ascribed to representatives of occupied site classes. The latter is
similar to the existing thermodynamic models of liquid
silicates~\cite{Gaskell, Mysen}, including the Quasi-Chemical Model
and its modifications~\cite{Blander,Pelton1,Pelton2}, where the
internal energy is assumed to be composed of SNNB energies.

A compromise between these two extremes, 383 and 6 energy
parameters, consists in applying the single site energy model
(\ref{iso_site_energy}) to nonoverlapping face-diagonal
configurations $(u,y)$, whilst ascribing different energies to
representatives of the 31 isotropy equivalence class of overlapping
configurations, which gives 6 + 31 = 37 energy parameters.

\subsection{Constrained minimization}


With $\cN_{\rm SiO_2}$ and $\cN_{\rm M_2O}$ denoting the mole numbers
of ${\rm SiO_2}$ and ${\rm M_2O}$, and
\begin{equation}
\label{x}
    x
    :=
    \frac
    {\cN_{\rm SiO_2}}
    {\cN_{\rm SiO_2} + \cN_{\rm M_2O}}
\end{equation}
denoting the ${\rm SiO_2}$ mole fraction, the centralised mole fractions of
Si and M atoms can be calculated by using (\ref{Ni/N1}) as
\begin{align}
\label{y2} y_2
    & =
    \frac
    {\cN_{\rm Si}}
    {\cN_{\rm O}}
    =
    \frac{x}{1+x},    \\
\label{y3} y_3
     & =
    \frac
    {\cN_{\rm M}}
    {\cN_{\rm O}}
    =
    \frac{2(1-x)}{1+x}
\end{align}
and satisfy the identity $
    2y_2 + \frac{1}{2}y_3
    =
    1
$. In accordance with Table~\ref{tab1}, the matrix $\Ups$, given
by (\ref{Ups}), takes the form
$$
    \Ups
    =
    \begin{bmatrix}
    12 &    4 &     24 &     24 &     12&      3\\
     6 &    2 &     6  &    6   &   0   &   0    \\
     0 &    0 &    24  &   24   &  24   &   6
    \end{bmatrix}
$$
and has rank two since its rows are linearly related as $\Ups_{1\bullet} =
2\Ups_{2\bullet} + \frac{1}{2}\Ups_{3\bullet}$.

The matrix $\wt{M}$ in (\ref{M}) turns out to be of size
$28 \x 383$ and full row rank. Therefore, the marginalization and
balance equations (\ref{thetaMDLS}) and (\ref{iso_balance_system}) lead to
28+2 = 30 independent linear equations for 383 + 6 = 389 variables.
The fact that the combined set of marginalization and
balance equations is highly under-determined favours the Newton
scheme of Section~\ref{sec:Lagrange} for this application which involves 30 Lagrange
multipliers instead of 389 dependent variables. However, the above mentioned
issues of finding a good initial approximation for the iterative algorithm and avoiding false extrema (which arise from the
non-convexity of the problem) require additional investigation.

\subsection{Gibbs energy of mixing}
    \label{sec:Gibbs_energy}

In view of (\ref{x}), the mixture of $0< x < 1$ moles of ${\rm
SiO_2}$ and $1- x$ moles of ${\rm M_2 O}$ contains $1+x$ moles of
oxygen atoms. Therefore, the approximation
$\wh{\bG}_{\face}(x)$ of the Gibbs energy $\bG(x)$ for the
liquid silicate can be  calculated in terms of (\ref{hatggg}) as
\begin{equation}
\nonumber
    \frac{\wh{\bG}_{\face}(x)}{RT}
     =
    (1+x)
    \wh{\bg}(x),
\end{equation}
where  $ R = 8.314$${\rm \frac{J}{mole\x K}}$ is the universal gas constant, and
we have indicated the dependence of $\wh{\bg}$
on the ${\rm SiO_2}$ mole fraction $x$ which enters the constrained minimization problem through the balance equations. Hence,
the corresponding approximation $\wh{\Delta \bG}_{\face}(x)$ of
the Gibbs energy of mixing
 can be
found from
\begin{equation}
\nonumber
    \frac
    {\wh{\Delta \bG}_{\face}(x)}
    {RT}
     =
(1+x)\wh{\bg}(x) - 2x\wh{\bg}(1) - (1-x)\wh{\bg}(0).
\end{equation}

\subsection{Mass density}
    \label{sec:mass_density}

In the framework of the face-to-diagonal reduction of the
Kramers-Wannier approximation, the mass density $\rho$ of the ${\rm
SiO_2}$--${\rm M_2O}$ liquid silicate with chemical composition
quantified by (\ref{x}) is estimated as
\begin{align*}
    \wh{\rho}_{\face}
    & =
    \frac
    {
        m_{\rm O} \cN_{\rm O}
        +
        m_{\rm Si} \cN_{\rm Si}
        +
        m_{\rm M} \cN_{\rm M}
    }
    {V_1 \cN }\\
    & =
    \big(
        1 -
        \wh{\cS}_0
    \big)
    \frac
    {
        (1+x) m_{\rm O}
        +
        x m_{\rm Si} + 2(1-x) m_{\rm M}
    }
    {(1+x) V_1},
\end{align*}
where $m_{\rm O}$, $m_{\rm Si}$ and $m_{\rm M}$ are the masses of O,
Si and M atoms, respectively,  and use is made of
(\ref{N1}), (\ref{y2}) and (\ref{y3}). As a result of the
blended minimization in (\ref{hatggg}),
the fraction
 $\wh{\cS}_0$ of oxygen-free sites of the carrier lattice depends not only on
the pressure and temperature through the quantity $K$ defined by
(\ref{K}) but also on the ${\rm SiO_2}$ mole fraction $x$. This
dependence reflects a subtle interplay between the geometric
constraints and energetics of the system.

\section{Conclusion}\label{sec:conclusion}


For the class of NNE-constrained interacting particle systems on the simple cubic lattice, which is relevant for statistical mechanical modeling of disordered condensed media, we have carried out a face-to-diagonal reduction of the Kramers-Wannier entropy density approximation. The latter represents the second level in the hierarchy of CVM approximations and takes into account the statistical correlations of the equilibrium state within faces of cubic cells.

Using a separation-of-variables technique in the framework of this entropy density approximation, we have obtained equations for approximate computation of the equilibrium Gibbs free energy in a class of lattice models of chemical systems with complex site configurations and short-range ordering, and a three-loop architecture for its numerical  implementation has been proposed.

We have outlined an application of the statistical mechanical approach to thermodynamic modeling of a binary liquid silicate formed from silica and the oxide of a univalent metal. The combinatorial and numeric aspects of the computer implementation of the model have been discussed. The results on model calibration for specific systems (such as sodium silicate ${\rm SiO_2}$--${\rm Na_2O}$) are postponed in view of additional work on a reliable algorithm required for solving the subsidiary non-convex minimization problem.

The NNE-constrained cubic lattice setting is amenable to further refinement in the form of  a ``quasi-tetrahedron'' reduction of Kikuchi's cube approximation of CVM for more subtle predictions of thermodynamic properties of such systems. The latter development will be communicated in subsequent publications.

\begin{acknowledgments}
This work was carried out in 2007--2008 while the author was with the University of Queensland. The work was supported by a UQ Research and Development grant and the Australian Research Council. Helpful discussions on chemical thermodynamics with Dmitry Saulov, Alex Kondratiev, Eugene Jak, Peter Hayes and Alex Klimenko are also gratefully acknowledged.
\end{acknowledgments}

\appendix

\section{Unique solvability of the decic equation}\label{sec:unique}

The decic equation (\ref{decic}) is equivalent to $\chi$ being a zero of
the function
$$
    f(x)
    :=
    x+
    \frac{1}{2}
    \Big(
    \re^{-K/3}
    g(\theta x)
    - 1
    \Big)
$$
on the interval $(0,1/2]$.
Here,
$$
    g(x)
    :=
    (1-2x)^3
    (1-x)^{-7/3}
$$
is a strictly convex function which decreases strictly from $g(0) = 1$ to $g(1/2) = 0$. The
monotonicity of $g$ follows from
\begin{eqnarray*}
    g'(x)
    & = &
    \frac{1}{3}
    (1-2x)^2
    (1-x)^{-10/3}
    (4x-11)
    < 0.
\end{eqnarray*}
Hence, for any $K
> 0$ and $\theta \in [1/2, 1)$, the function $f$
satisfies
 \begin{eqnarray*}
    f(0)
    =
    \frac{\re^{-K/3} - 1}{2}
    < 0 <
    \frac{\re^{-K/3}g(\theta/2)}{2}
    = f(1/2),
 \end{eqnarray*}
which, by the Intermediate  Value Theorem, implies that the equation
$f(x) = 0$ has a root on $(0,1/2]$. The uniqueness of the root follows from the
convexity of $f$ inherited from $g$. The latter is established by
\begin{eqnarray*}
    g''(x)
    & = &
    \frac{2}{9}
    (1-2x)
    (1-x)^{-13/3}
    (-4x^2 + 22x + 17) \> 0,
\end{eqnarray*}
where both roots of the rightmost quadratic polynomial are beyond the interval
 $(0, 1/2]$.

\section{Newton iterations}
\label{sec:newton}

The constrained optimization problem of Section~\ref{sec:Lagrange}
is a particular case of the minimization problem
$$
    f(p)
    :=
    \bra a, p\ket
    +
    \bra b, \Lambda(p)\ket
    \longrightarrow
    \min,
$$
where the minimum is taken over a finite-dimensional column-vector
$p$ of probabilities
subject to the system of linear constraints
$$
    Ap = q.
$$
Here, the function $\Lambda$,  defined by (\ref{Lambda}), applies
entry-wise; $a$, $b$ and $q$ are appropriately dimensioned
column-vectors, with $b$ consisting of nonzero entries, and the matrix $A$
is of full row rank. The linear part $\bra a, p\ket$ of the
objective function is the internal energy term, while $\bra b,
\Lambda(p)\ket$ originates from an entropy cumulant expression,
typical for CVM.
If the entries of $b$ are all positive, then the function $f$ is strictly convex.
However, if some of them are negative, as is the case in the
above mentioned problem of Section~\ref{sec:Lagrange}, the overall
convexity of $f$ under the linear constraints is a nontrivial issue  \cite{Pelizzola}
 which complicates the numerical solution of this
problem. At the level of first-order necessary conditions of
optimality for relatively interior points, the stationarity of the
Lagrange function
$$
    \cL(p,\tau)
    :=
    f(p)
    -
    \bra
        \tau,
        Ap-q
    \ket
$$
with respect to $p$ is  equivalent to that the gradient of $f$ satisfies
$$
    f'(p) = A^{\rT} \tau,
$$
which allows $p$ to be expressed in terms of the column-vector
$\tau$ of Lagrange multipliers  as
$$
    \wh{p}
    :=
    \exp
    \left(
        (
            A^{\rT}\tau - a
        )/b
        -\bone
    \right).
$$
Here, the exponential function and division are applied to vectors entry-wise, and
$(\cdot)^{\rT}$ denotes the matrix transpose.
The Jacobian matrix of the map $\wh{p}$ with respect to $\tau$
is computed as
$$
    \wh{p}\,'
    =
    (f''(\wh{p}))^{-1} A^{\rT}
    =
    \diagmat(\wh{p} /b) A^{\rT},
$$
where
$$
    f''(p) = \diagmat(b / p)
$$
is the Hessian  matrix of the function $f$, and $\diagmat(v)$ is the
diagonal matrix with the vector $v$ over the main diagonal. Hence, in order to satisfy the
linear constraints, the Newton iterations for $\tau$ take the form
$$
    \tau
    \mapsto
    \tau
    -
    \left(
        A
        \diagmat(\wh{p} / b)
        A^{\rT}
    \right)^{-1}
    \left(
        A\wh{p}-q
    \right).
$$
Here, the invertibility of the matrix is not guaranteed if $b$
contains both positive and negative entries, even though $A$ is of
full row rank.
By assuming that the iterates converge to $\wh{\tau}$ and using
the identity
$$
    \bra
        p,
        f'(p)
    \ket
    =
    f(p)
    +
    \bra
        b,
        p
    \ket,
$$
it follows that the corresponding value of the function is calculated as
$$
    f(\wh{p})
    =
    \bra
        q,
        \wh{\tau}
    \ket
    -
    \bra
        b,
        \wh{p}
    \ket.
$$
In the complicated case, where some entries of the vector $b$ are negative, a
sufficient condition for the point $\wh{p}$ to be a local minimum under
the linear constraints is the positive definiteness of the Hessian
matrix $f''(\wh{p})$ on the null space of  the matrix $A$.


\bibliography{CVM_NNE_Kramers_Wannier_final_cut_arxiv}

\begin{thebibliography}{23}
\expandafter\ifx\csname natexlab\endcsname\relax\def\natexlab#1{#1}\fi
\expandafter\ifx\csname bibnamefont\endcsname\relax
  \def\bibnamefont#1{#1}\fi
\expandafter\ifx\csname bibfnamefont\endcsname\relax
  \def\bibfnamefont#1{#1}\fi
\expandafter\ifx\csname citenamefont\endcsname\relax
  \def\citenamefont#1{#1}\fi
\expandafter\ifx\csname url\endcsname\relax
  \def\url#1{\texttt{#1}}\fi
\expandafter\ifx\csname urlprefix\endcsname\relax\def\urlprefix{URL }\fi
\providecommand{\bibinfo}[2]{#2}
\providecommand{\eprint}[2][]{\url{#2}}

\bibitem[{\citenamefont{Vladimirov and Jak}(2007)}]{Vladimirov}
\bibinfo{author}{\bibfnamefont{I.}~\bibnamefont{Vladimirov}} \bibnamefont{and}
  \bibinfo{author}{\bibfnamefont{E.}~\bibnamefont{Jak}}, \bibinfo{journal}{J.
  Chem. Phys.} \textbf{\bibinfo{volume}{126}}, \bibinfo{pages}{164502}
  (\bibinfo{year}{2007}).

\bibitem[{\citenamefont{Verhagen}(1977)}]{Verhagen}
\bibinfo{author}{\bibfnamefont{A.}~\bibnamefont{Verhagen}},
  \bibinfo{journal}{J. Chem. Phys.} \textbf{\bibinfo{volume}{67}},
  \bibinfo{pages}{5060} (\bibinfo{year}{1977}).

\bibitem[{\citenamefont{Baxter}(1999)}]{Baxter}
\bibinfo{author}{\bibfnamefont{R.}~\bibnamefont{Baxter}},
  \bibinfo{journal}{Annals of Combinatorics} \textbf{\bibinfo{volume}{3}},
  \bibinfo{pages}{191} (\bibinfo{year}{1999}).

\bibitem[{\citenamefont{Huang}(1987)}]{Huang}
\bibinfo{author}{\bibfnamefont{K.}~\bibnamefont{Huang}},
  \emph{\bibinfo{title}{Statistical Mechanics}} (\bibinfo{publisher}{John Wiley
  \& Sons}, \bibinfo{address}{New York}, \bibinfo{year}{1987}),
  \bibinfo{edition}{2nd} ed.

\bibitem[{\citenamefont{An}(1988)}]{An}
\bibinfo{author}{\bibfnamefont{G.}~\bibnamefont{An}}, \bibinfo{journal}{J.
  Statist. Phys.} \textbf{\bibinfo{volume}{52}}, \bibinfo{pages}{727}
  (\bibinfo{year}{1988}).

\bibitem[{\citenamefont{Kikuchi}(1951{\natexlab{a}})}]{Kikuchi1}
\bibinfo{author}{\bibfnamefont{R.}~\bibnamefont{Kikuchi}},
  \bibinfo{journal}{Phys. Rev.} \textbf{\bibinfo{volume}{81}},
  \bibinfo{pages}{988} (\bibinfo{year}{1951}{\natexlab{a}}).

\bibitem[{\citenamefont{Kikuchi}(1951{\natexlab{b}})}]{Kikuchi2}
\bibinfo{author}{\bibfnamefont{R.}~\bibnamefont{Kikuchi}}, \bibinfo{journal}{J.
  Chem. Phys.} \textbf{\bibinfo{volume}{19}}, \bibinfo{pages}{1230}
  (\bibinfo{year}{1951}{\natexlab{b}}).

\bibitem[{\citenamefont{Kurata et~al.}(1953)\citenamefont{Kurata, Kikuchi, and
  Watari}}]{Kikuchi3}
\bibinfo{author}{\bibfnamefont{M.}~\bibnamefont{Kurata}},
  \bibinfo{author}{\bibfnamefont{R.}~\bibnamefont{Kikuchi}}, \bibnamefont{and}
  \bibinfo{author}{\bibfnamefont{T.}~\bibnamefont{Watari}},
  \bibinfo{journal}{J. Chem. Phys.} \textbf{\bibinfo{volume}{21}},
  \bibinfo{pages}{434} (\bibinfo{year}{1953}).

\bibitem[{\citenamefont{Kikuchi and Masuda-Jindo}(2002)}]{Kikuchi_2002}
\bibinfo{author}{\bibfnamefont{R.}~\bibnamefont{Kikuchi}} \bibnamefont{and}
  \bibinfo{author}{\bibfnamefont{K.}~\bibnamefont{Masuda-Jindo}},
  \bibinfo{journal}{Calphad} \textbf{\bibinfo{volume}{26}}, \bibinfo{pages}{33}
  (\bibinfo{year}{2002}).

\bibitem[{\citenamefont{Moran-Lopez and Sanchez}(1996)}]{Moran}
\bibinfo{editor}{\bibfnamefont{J.}~\bibnamefont{Moran-Lopez}} \bibnamefont{and}
  \bibinfo{editor}{\bibfnamefont{J.}~\bibnamefont{Sanchez}}, eds.,
  \emph{\bibinfo{title}{Theory and Applications of the Cluster Variation and
  Path Probability Methods}} (\bibinfo{publisher}{Plenum Press},
  \bibinfo{address}{New York}, \bibinfo{year}{1996}).

\bibitem[{\citenamefont{Pelizzola}(2005)}]{Pelizzola}
\bibinfo{author}{\bibfnamefont{A.}~\bibnamefont{Pelizzola}},
  \bibinfo{journal}{J. Phys. A: Math. Gen.} \textbf{\bibinfo{volume}{38}},
  \bibinfo{pages}{R309} (\bibinfo{year}{2005}).

\bibitem[{\citenamefont{Cover and Thomas}(2006)}]{Cover}
\bibinfo{author}{\bibfnamefont{T.}~\bibnamefont{Cover}} \bibnamefont{and}
  \bibinfo{author}{\bibfnamefont{J.}~\bibnamefont{Thomas}},
  \emph{\bibinfo{title}{Elements of Information Theory}}
  (\bibinfo{publisher}{Wiley}, \bibinfo{address}{Hoboken, New Jersey},
  \bibinfo{year}{2006}), \bibinfo{edition}{2nd} ed.

\bibitem[{\citenamefont{Cotton}(1990)}]{Cotton}
\bibinfo{author}{\bibfnamefont{F.}~\bibnamefont{Cotton}},
  \emph{\bibinfo{title}{Chemical Applications of Group Theory}}
  (\bibinfo{publisher}{Wiley}, \bibinfo{address}{New York},
  \bibinfo{year}{1990}), \bibinfo{edition}{3rd} ed.

\bibitem[{\citenamefont{Enderton}(2001)}]{Enderton}
\bibinfo{author}{\bibfnamefont{H.}~\bibnamefont{Enderton}},
  \emph{\bibinfo{title}{A Mathematical Introduction to Logic}}
  (\bibinfo{publisher}{Harcourt Academic Press, San Diego},
  \bibinfo{year}{2001}), \bibinfo{edition}{2nd} ed.

\bibitem[{\citenamefont{Gaskell}(1981)}]{Gaskell}
\bibinfo{author}{\bibfnamefont{D.}~\bibnamefont{Gaskell}},
  \bibinfo{journal}{Canad. Met. Quaterly} \textbf{\bibinfo{volume}{20}},
  \bibinfo{pages}{3} (\bibinfo{year}{1981}).

\bibitem[{\citenamefont{Mysen}(1988)}]{Mysen}
\bibinfo{author}{\bibfnamefont{B.}~\bibnamefont{Mysen}},
  \emph{\bibinfo{title}{Structure and Properties of Silicate Melts}}
  (\bibinfo{publisher}{Elsevier Science Publishers},
  \bibinfo{address}{Amsterdam}, \bibinfo{year}{1988}).

\bibitem[{\citenamefont{Blander and Pelton}(1984)}]{Blander}
\bibinfo{author}{\bibfnamefont{M.}~\bibnamefont{Blander}} \bibnamefont{and}
  \bibinfo{author}{\bibfnamefont{A.~D.} \bibnamefont{Pelton}}, in
  \emph{\bibinfo{booktitle}{Proc. 2nd Int. Symp. Met. Slags and Fluxes}}
  (\bibinfo{organization}{TMS--AIME}, \bibinfo{address}{Warrendale, PA},
  \bibinfo{year}{1984}), pp. \bibinfo{pages}{295--304}.

\bibitem[{\citenamefont{Pelton and Blander}(1986)}]{Pelton1}
\bibinfo{author}{\bibfnamefont{A.}~\bibnamefont{Pelton}} \bibnamefont{and}
  \bibinfo{author}{\bibfnamefont{M.}~\bibnamefont{Blander}},
  \bibinfo{journal}{Met. Trans. B} \textbf{\bibinfo{volume}{17B}},
  \bibinfo{pages}{805} (\bibinfo{year}{1986}).

\bibitem[{\citenamefont{Pelton et~al.}(2000)\citenamefont{Pelton, Degterov,
  Eriksson, Robelin, and Dessureault}}]{Pelton2}
\bibinfo{author}{\bibfnamefont{A.}~\bibnamefont{Pelton}},
  \bibinfo{author}{\bibfnamefont{S.}~\bibnamefont{Degterov}},
  \bibinfo{author}{\bibfnamefont{G.}~\bibnamefont{Eriksson}},
  \bibinfo{author}{\bibfnamefont{C.}~\bibnamefont{Robelin}}, \bibnamefont{and}
  \bibinfo{author}{\bibfnamefont{Y.}~\bibnamefont{Dessureault}},
  \bibinfo{journal}{Met. Mat. Trans. B} \textbf{\bibinfo{volume}{31B}},
  \bibinfo{pages}{651} (\bibinfo{year}{2000}).

\bibitem[{\citenamefont{Kindermann and Snell}(1980)}]{Kindermann}
\bibinfo{author}{\bibfnamefont{R.}~\bibnamefont{Kindermann}} \bibnamefont{and}
  \bibinfo{author}{\bibfnamefont{J.}~\bibnamefont{Snell}},
  \emph{\bibinfo{title}{Markov Random Fields and Their Applications}}
  (\bibinfo{publisher}{American Mathematical Society},
  \bibinfo{address}{Providence}, \bibinfo{year}{1980}).

\bibitem[{\citenamefont{Preston}(1976)}]{Preston}
\bibinfo{author}{\bibfnamefont{C.}~\bibnamefont{Preston}},
  \emph{\bibinfo{title}{Random Fields}} (\bibinfo{publisher}{Springer-Verlag},
  \bibinfo{address}{Berlin}, \bibinfo{year}{1976}).

\bibitem[{\citenamefont{Conway and Sloane}(1988)}]{Conway}
\bibinfo{author}{\bibfnamefont{J.}~\bibnamefont{Conway}} \bibnamefont{and}
  \bibinfo{author}{\bibfnamefont{N.}~\bibnamefont{Sloane}},
  \emph{\bibinfo{title}{Sphere Packings, Lattices and Groups}}
  (\bibinfo{publisher}{Springer-Verlag}, \bibinfo{address}{New York},
  \bibinfo{year}{1988}).

\bibitem[{\citenamefont{Shiryaev}(1995)}]{Shiryaev}
\bibinfo{author}{\bibfnamefont{A.}~\bibnamefont{Shiryaev}},
  \emph{\bibinfo{title}{Probability}} (\bibinfo{publisher}{Springer},
  \bibinfo{address}{Berlin}, \bibinfo{year}{1995}), \bibinfo{edition}{2nd} ed.

\end{thebibliography}

\end{document}